%% file: manuscript.tex
\documentclass[%
 reprint,
superscriptaddress,
 amsmath,amssymb,
 aps,prl
]{revtex4-1}

\usepackage{dsfont}
\usepackage{graphicx}
\usepackage{dcolumn}
\usepackage{bm}
\usepackage[dvipsnames]{xcolor}
\usepackage{hyperref}
\hypersetup{
colorlinks=true,
	linkcolor=BlueViolet,
	citecolor=BlueViolet,
	urlcolor=BlueViolet,
}

\usepackage{ulem} 

\newcommand{\comment}[1]{}

\begin{document}
\definecolor{nrppurple}{RGB}{128,0,128}

\preprint{APS/123-QED}

\title{A transmon qubit realized by exploiting the superconductor-insulator transition}

\author{C.~G.~L.~B\o{}ttcher}
\thanks{Email: charlotte.boettcher@stanford.edu, and \\
michel.devoret@yale.edu.}
\affiliation{Department of Applied Physics, Yale University, New Haven, CT 06520, USA}
\affiliation{Department of Applied Physics, Stanford University, Stanford, CA 94305, USA}
\author{E.~{\"O}nder}
\affiliation{Department of Applied Physics, Yale University, New Haven, CT 06520, USA}
\author{T.~Connolly}
\affiliation{Department of Applied Physics, Yale University, New Haven, CT 06520, USA}
\author{J.~Zhao}
\affiliation{Center for Quantum Devices, Niels Bohr Institute, University of Copenhagen, 2100 Copenhagen, Denmark}
\author{C.~Kvande}
\affiliation{Center for Quantum Devices, Niels Bohr Institute, University of Copenhagen, 2100 Copenhagen, Denmark}
\affiliation{Department of Physics, University of Washington, Seattle, Washington 98195, USA}
\author{D.~Q.~Wang}
\affiliation{Department of Electrical Engineering, Yale University, New Haven, CT 06520, USA}
\author{P.~D.~Kurilovich}
\affiliation{Department of Physics, Yale University, New Haven, CT 06520, USA}
\affiliation{Department of Applied Physics, Yale University, New Haven, CT 06520, USA}
\author{S.~Vaitiek\.{e}nas}
\affiliation{Center for Quantum Devices, Niels Bohr Institute, University of Copenhagen, 2100 Copenhagen, Denmark}
\author{L.~I.~Glazman}
\affiliation{Department of Physics, Yale University, New Haven, CT 06520, USA}
\affiliation{Yale Quantum Institute, Yale University, New Haven, Connecticut 06511, USA}
\author{H.~X.~Tang}
\affiliation{Department of Electrical Engineering, Yale University, New Haven, CT 06520, USA}
\author{M.~H.~Devoret}
\thanks{Email: charlotte.boettcher@stanford.edu, and \\
michel.devoret@yale.edu.}
\affiliation{Department of Applied Physics, Yale University, New Haven, CT 06520, USA}
\affiliation{Department of Physics, UCSB, Santa Barbara, CA 93106, USA}
\affiliation{Google Quantum AI, 301 Mentor Dr, Goleta, California 93111, USA}

\date{\today}

\begin{abstract}
    Superconducting qubits are among the most promising platforms for realizing practical quantum computers. One requirement to create a quantum processor is nonlinearity, which in superconducting circuits is typically achieved by sandwiching a layer of aluminum oxide between two aluminum electrodes to form a Josephson junction. These junctions, however, face several limitations that hinder their scalability: the small superconducting gap of aluminum necessitates millikelvin operating temperatures, the material interfaces lead to dissipation, and the sandwich geometry adds unwelcome capacitance for high-frequency applications. In this work, we address all three limitations using a novel superconducting weak link based on the superconductor-insulator transition. By locally thinning a single film of niobium nitride, we exploit its thickness-driven superconductor-insulator transition to form a weak link employing only atomic layer deposition and atomic layer etching. We utilize our weak links to produce a transmon qubit, '\textit{planaron}', with a measured anharmonicity of $\alpha/2\pi = 235$ MHz; at present, the linewidth is $\kappa/2\pi = 15 \mathrm{\: MHz}$. The high superconducting gap of niobium nitride can enable operation at elevated temperatures in future devices, and the fully planar geometry of the weak link  eliminates superfluous material interfaces and capacitances. The investigation of small patches of material near the SIT can shed new light on the nature of the transition, including the role of dissipation and finite-size effects.
    
    
   
\end{abstract}

\maketitle
Building a scalable quantum computer is thought by many as climbing the Everest of quantum technology. Superconducting qubits are one of the main contenders in solid-state platforms for quantum computing \cite{Nakamura1999-la,Google-Quantum-AI-and-Collaborators2025-ct}. However, the requirement for millikelvin operating temperatures poses a significant challenge for scalability. Superconducting qubits that can function at temperatures above 1~K could eliminate the requirement for dilution refrigerators, making large-scale devices more feasible \cite{Krinner2019-ps,Martin2022-ge}. A main challenge in achieving higher operating temperatures stems from the aluminum-based Josephson junction used in today's superconducting qubits. Its low superconducting gap restricts operations to below $\sim 200 \mathrm{\ mK}$ due to decoherence from quasiparticles in thermal equilibrium~\cite{glazman2021}. Typical metal oxide sandwich junctions also suffer from parasitic capacitance, which limits operating frequencies to below the plasma frequency of the junction $\omega_p = 1/\sqrt{L_JC_J}$, where $L_J$ and $C_J$ are junction inductance and capacitance, respectively. This is problematic for high-temperature qubits, since they must operate at frequencies much higher than the temperature to mitigate thermal excitations. As the operating frequency approaches the plasma frequency, the participation of the lossy oxide barrier increases, reducing coherence times.

Recent efforts have addressed some of these limitations. One approach is to proximitize aluminum junctions with large-gap superconductors such as niobium, enabling qubit operation near 1 K~\cite{anferov2024,anferov2025}. Another technique is to implement fully nitride-based architectures, such as NbN/AlN/NbN tunnel junctions~\cite{wang2025} or TiN nanowires \cite{purmessur2025}.
\newline
\newline
\indent In this work, while remaining within the framework of a nitride-based architecture, we pursue an alternative novel approach. We eliminate the multilayer oxide junction and introduce a monolithic weak link in a single film of niobium nitride (NbN), a high-gap superconductor. Our device leverages the superconductor-insulator transition (SIT), a quantum phase transition that can be tuned by film thickness. Using a combination of atomic layer deposition (ALD) and atomic layer etching (ALE), we locally thin a uniform NbN film near the SIT and define a weak link in a fully planar geometry.
The atomically defined link contrasts with other techniques such as focused ion beam (FIB) etching \cite{troeman2007,MATSUMOTO2011,couedo2020,ruhtinas2023}, known to induce structural disorder~\cite{SALVATI2018,shani2022}.

The superconductor-insulator transition (SIT) belongs to the class of fundamental quantum phase transitions displayed by two-dimensional (2D) superconducting systems \cite{Fisher1990,Goldman2010}. It has been extensively studied across a variety of systems ranging from amorphous \cite{Lee1990,steiner2008} and granular \cite{Jaeger1989,bollinger2011} thin films to two-dimensional Josephson junction arrays \cite{eley2012,han2014,Boettcher2018,Sasmal2025}. 
This transition is typically tuned by adjusting parameters such as disorder, magnetic field, or electron density.
On the superconducting side of the transition, the phase of the local order parameter is a good quantum number, and Cooper pairs flow without resistance; while on the insulating side the conjugate variable, charge, tends to be a good quantum number, and it is instead vortices that flow freely. Superconducting qubits are circuits with only one degree of freedom, and they too present regimes where either charge or phase is a good quantum number. We realize a superconducting qubit from a patch of material that is tuned close to its SIT, and explore a range of patch parameters that allow us to build weak links with highly tunable properties. The main result of this exploration is the exploitation of an ALE-driven SIT in NbN to realize a transmon qubit with a measured anharmonicity of 235 MHz.\\
\indent This work further introduces a new system to investigate the nature of the superconductor-insulator transition which is governed by a complex interplay between disorder, superconducting pairing, and Coulomb repulsion. This interplay is still not fully understood in practical materials, and the investigation of mesoscopic samples with parameters close to the SIT transition may shed new light on the nature of the transition. 
Remarkably, an even simpler theory of Josephson tunneling across a small patch of disordered conductor close to the Anderson localization transition, which does not account for electron repulsion, remains an open question~\cite{asano2002}. The dual goal of this work is to investigate the Josephson effect near the SIT transition and to harness it for realizing a new type of qubit. We compare our experimental results for the weak link behavior against two theoretical models: that of a diffusive metal weak link and that of the Anderson insulator weak link. We find that our system exhibits features of both regimes. Some of our results – such as the temperature dependence of the critical current – are more consistent with the diffusive model, while others – such as the small extracted value of the electron mean free path – are more consistent with the localization model. This suggests that our weak links are in the borderline regime between the two cases.

\begin{figure}[t]
	\includegraphics[width=3.6 in]{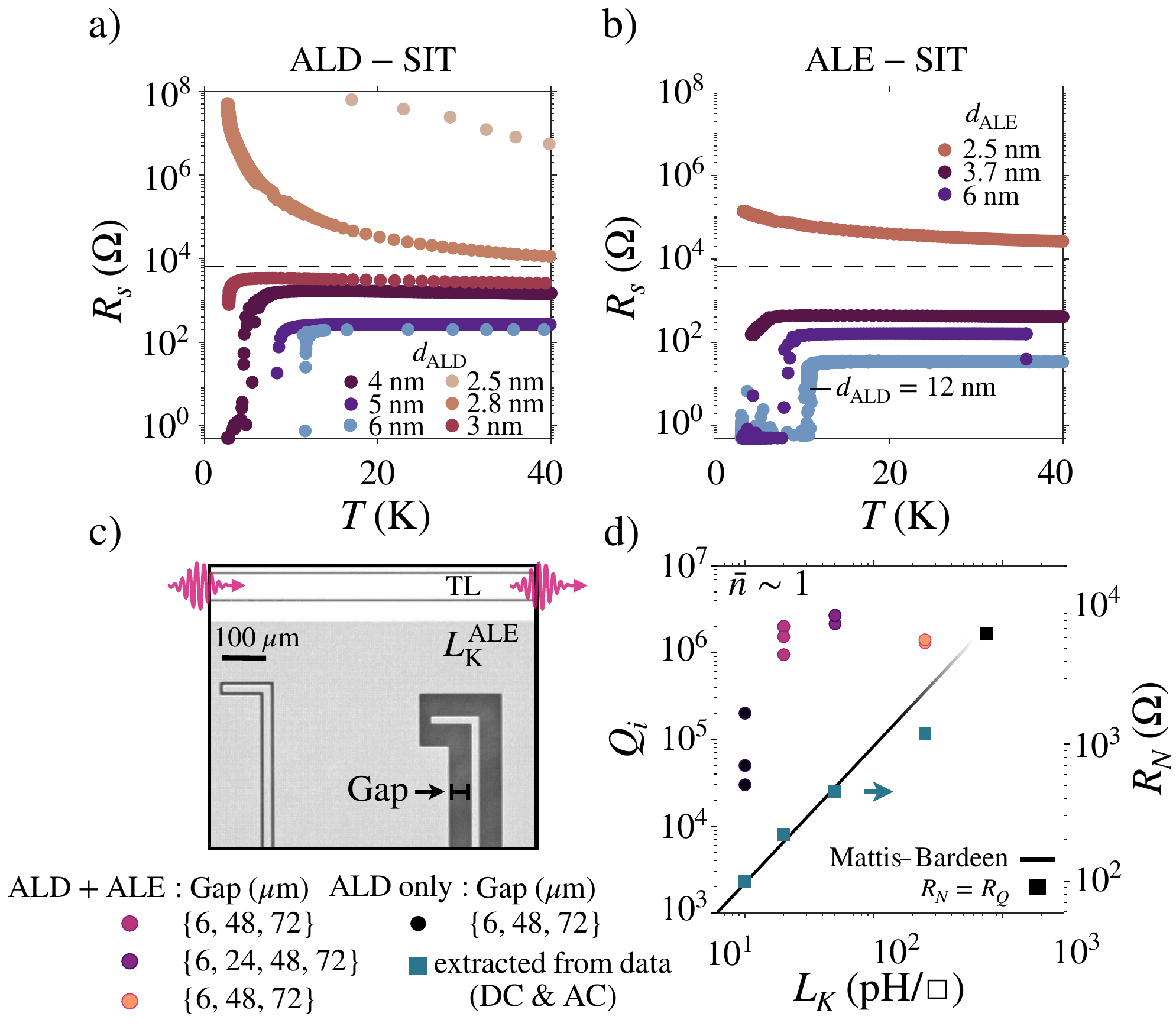}
	\caption{\textbf{ALD- and ALE-driven SIT in NbN thin films} DC transport experimental data on the thickness-driven SIT in NbN films obtained by a) atomic layer deposition (ALD) and b) atomic layer etching (ALE). The two panels show that film resistance per square, $R_s$, varies as a function of thickness, $d_{\mathrm{ALD}}$ and $d_{\mathrm{ALE}}$, and temperature. A separation between a superconducting ($dR_s(T)/dT > 0$) and an insulating ($dR_s(T)/dT < 0$) behavior is observed for an ALD-driven SIT around $d^{\mathrm{ALD}}_c\sim 3$ nm. Horizontal dashed lines in a,b) mark a sheet resistance per square of $R_Q=6.45\rm \:k\Omega$. Within experimental uncertainties, the ALE-driven transition occur at a similar thickness. c) Optical image of a representative CPW resonator chip with locally tuned thickness. The image shows two out of four resonators capacitively coupled to a transmission line (TL) with varying gaps between center conductor and ground plane ($\mathrm{gap = \{6,\:24,\:48,\:72\}~\mu m}$). Dark grey is the sapphire substrate, light grey is ALE-thinned NbN containing the CPW resonators, and white is the unetched film containing the transmission line. d) \textit{Left axis:} Measured quality factors of NbN CPW resonators at single photon power. Quality factors of $\sim10^6$ are obtained for ALE etched films (pink, purple, orange circles), independent of both kinetic inductance and the gap between the center conductor and the ground plane. Unetched films (black circles) show a reduced quality factor which depends on the gap (see Supplemental Information for discussion). \textit{Right axis:} The normal-state sheet resistance, $R_N$, for each film is obtained from DC Hall bar measurements on the same chip and plotted as a function of the kinetic inductance, $L_K$, extracted from CPW measurements (teal squares). The relation is in good agreement with Mattis-Bardeen (black solid line). Black square marks Mattis-Bardeen prediction when $R_N=R_Q$.
			}
	\label{fig1}

\end{figure}
\section{I. Thickness-driven SIT in niobium nitride}
We first investigate the thickness-controlled SIT in uniform 2D films of NbN.
We can define precise thicknesses using either the additive process of atomic layer deposition [ALD, Fig.~\ref{fig1}a] or the subtractive process of atomic layer etching [ALE, Fig.~\ref{fig1}b], after the growth of a thick film by ALD. The low-temperature sheet resistance per square, $R_s$, varies by several orders of magnitude over a narrow range of thicknesses, indicating a critical thickness, $d_c\sim 2.5-3\:\mathrm{nm}$ that separates the superconducting ($dR_s(T)/dT > 0$) and the insulating ($dR_s(T)/dT < 0$) regimes. At the critical thickness, the normal-state sheet resistance per square, $R_{N}$, is close to the expectation $R_N\sim R_Q=h/4e^2$ [dashed lines in Figs.~\ref{fig1}a,b] for a Bose-dominated SIT \cite{Fisher1990,liu1991}. Neither the critical resistance nor the critical thickness depend significantly on whether we use ALD or ALE to achieve the final thickness.
\newline

\begin{figure}[t]
	\includegraphics[width=3.3 in]{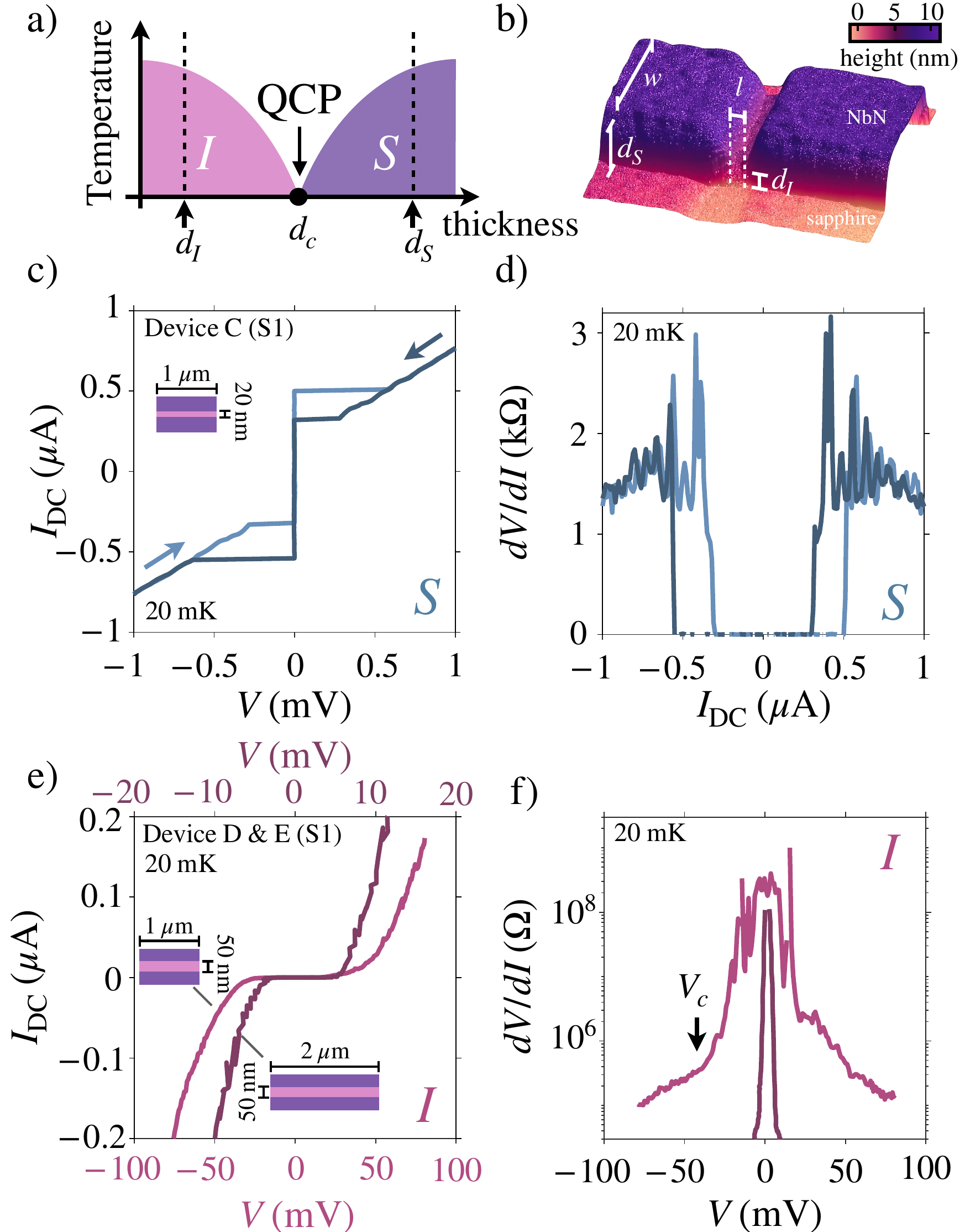}
	\caption{\textbf{Single-film planar weak links} a) Schematic phase diagram of the superconductor-insulator transition (SIT) tuned by film thickness. The quantum critical point (QPC) is marked at the critical thickness, $d_c$, that separates superconducting, $S$, from insulating, $I$, behavior near zero temperature. b) Atomic force microscopy (AFM) of a single-film weak link fabricated from a thick ALD-NbN film with a starting thickness of $d_S$, corresponding to a point deep in the superconducting phase, see panel a). Next, a local ALE etch defines the weak link with a thickness, $d_I$ of $\sim 2.5\:\rm nm$ corresponding to a point deep in the insulating phase. c, d) The $I-V$characteristic of link with $l = 20 \mathrm{\: nm}$. For sufficiently short links, we observe a supercurrent branch. e, f) The $I-V$ characteristics of two longer weak links, both with $l = 50 \mathrm{\: nm}$. As the length is increased, the supercurrent branch disappears and no current flows until a critical voltage $V_c\gg2\Delta_{\rm NbN}/e$ is reached.
    }
	\label{fig2}
\end{figure}

We next consider the possibility of a commonly observed, yet not fully understood, low-temperature metallic phase that may emerge between the superconducting and insulating phase in two dimensions, characterized by
the saturation of sheet resistance to a finite value as temperature approaches zero ($dR_s(T)/dT = 0$). Although the existence and nature of this phase remains under debate — whether it represents a genuine metallic ground state \cite{Kapitulnik2019,FEIGELMAN1998107,Diamantini2024} or arises from extreme sensitivity to environmental noise near the SIT \cite{tamir2019} — both scenarios may carry substantial implications for device coherence. As shown in Figs.~\ref{fig1}a and \ref{fig1}b, no anomalous metallic phase is observed in the present system with either ALD or ALE processing. To further explore a possible intervening metallic phase near the SIT beyond what might be visible in DC transport, we have also investigated dissipation in coplanar waveguide (CPW) resonators fabricated from films of varied thickness (tuned by ALE) near and far from the SIT, reflected in $R_N$. As shown in Fig.~\ref{fig1}d, internal quality factors in the single-photon regime, measured along with DC Hall bar transport on the same chip, exceed $10^6$ at base temperature ($\sim23\mathrm{\:mK}$), independent of $R_N$ in the range of investigation. If we view a disordered material as a network of Josephson junctions we can relate the impedance per square of the array to that of a single junction, resulting in the appearance of the Mattis-Bardeen relation $L_K=\hbar R_N/\pi \Delta$. We find that the extracted values of kinetic inductance, $L_K$, and the $R_N$ measured at DC follow closely the Mattis-Bardeen relation using $\Delta_\mathrm{NbN}=2.03$ meV (black solid line in Fig. \ref{fig1}d). We note, however, some deviation from the Mattis-Bardeen line near the SIT.
\newline
\indent The observation that $Q_i$ does not degrade with increased sheet inductance (i.e. proximity to the SIT) indicates the absence of both an increase in poisonous quasiparticles or a low-temperature metal phase. The uncorrupted $Q_i$ further indicates that the ALE process introduces minimal damage to the films. Notably, the quality factor of the resonators which undergo ALE is also independent of the gap between the center conductor and ground plane. The gap primarily sets the partition of interfaces, suggesting that the dissipation is not dominated by lossy material interfaces. The ALE process is gentle and reduces the surface roughness of the film (see Supplemental Information). This contrasts with reactive ion etching (RIE) or focused ion beam (FIB) etching, which introduces a greater degree of structural disorder, defects, and surface roughness. The improved surface quality of the ALE films may be related to the reduction in dissipation \cite{Mahuli2025,gerritsen2022}.
We have used bulk films for these initial film characterizations to calibrate the NbN thicknesses and corresponding electronic phases. The films here were large enough to use the SIT classification, which is defined in the infinite-size limit. Electron transport across small-size patches of the material with parameters in the vicinity of SIT will be addressed next.

\section{II. Single-film planar weak links}
To fabricate the planar weak links, we begin by depositing a uniform NbN film corresponding to the superconducting phase ($S$) on a sapphire wafer using ALD. The device structure is then defined through two sequential steps of local ALE. In the first step, e-beam lithography and ALE are used to define the weak link region by partially etching a long, narrow strip of NbN down to a target thickness corresponding to the insulating ($I$) phase [Fig.~\ref{fig2}a]. The narrow dimension of the strip defines the length, $l$, of the weak link. In the second step, we use e-beam lithography to define the width, $w$, of the weak link and the surrounding lead geometry. In this step, the NbN is completely removed in the selected areas using ALE [Fig.~\ref{fig2}b]. This fabrication process avoids the need for lift-off and is compatible with buffered oxide etch (BOE) cleaning. The subtractive method eliminates the possibility of trapping residual lithography resist between the film and the substrate and BOE cleaning enables robust elimination of oxides, organics, and fluoride contaminants for improved device quality.

The sequence described above is used to fabricate samples hereinafter denoted S1. We also fabricated samples for which the etching sequence is reversed. We will refer to these samples as S2 (see Supplemental Information for fabrication details).  We measured devices with several values of $w$ and $l$ for each fabrication procedure. We find that the qualitative behavior of the weak links is similar for both fabrication methods, as detailed further in the Supplemental Information.
\begin{figure*}[t]
	\includegraphics[width= 7 in]{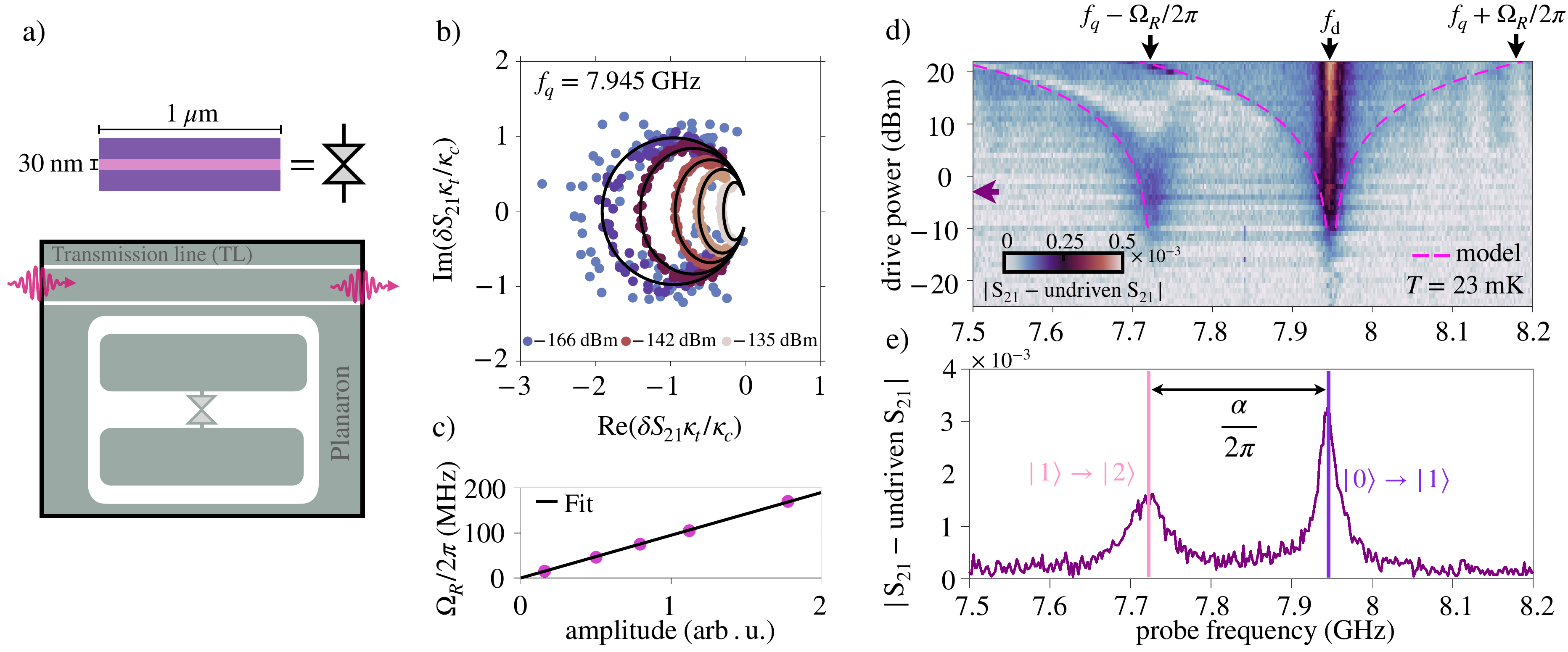}
	\caption{\textbf{Planaron qubit} a) \textit{Top:} Geometry of the superconducting junction type, represented with a butterfly symbol. \textit{Bottom:} The weak link is shunted with a large capacitor and directly coupled to a transmission line in a hanger-style geometry. b) Measured change in complex transmission coefficient, $\delta S_{21}$, as a function of drive power on the planaron qubit. Note that we subtract the high-power response of the system (listed applied powers account for total attenuation of the setup, which is calibrated using data in Fig. \ref{fig3}d). We also scale the measurement by $\kappa_t/\kappa_c$, such that the radius of the circle is unity at low drive powers. c) We extract the Rabi frequency, $\Omega_R$, using Eq.~(\ref{squash}) and plot it as a function of the drive amplitude. The Rabi rate is linear in drive amplitude as expected (black line). d) Two-tone spectroscopy of the planaron qubit. A drive is applied on resonance with the $|0\rangle \rightarrow|1\rangle$ transition. The frequency of a second probe tone is then swept to search for the $|1\rangle \rightarrow|2\rangle$ transition. The $S_{21}$ of the undriven planaron is subtracted from the results to make the resonance clearer. This subtraction results in a brightening of the $|0\rangle \rightarrow|1\rangle$ transition with drive power as the phase roll disappears [see main text]. At moderately high power, a second phase roll appears signifying the $|1\rangle \rightarrow|2\rangle$ transition. Sidebands are also observed at high drive powers due to the Aulter-Townes effect. They are fitted to the expected model (dashed pink lines) as described in main text. e) A line cut at $P_{RF}=-3 \:\mathrm{dBm}$ (-138 dBm at the device) displays two clear peaks $\sim 235$ MHz apart quantifying the anharmonicity, $\alpha/2\pi$, of the planaron qubit.
			}
	\label{fig3}
\end{figure*}

Figure \ref{fig2}c shows the current-voltage ($I\mathrm{\tiny-}V$) characteristics of a representative weak link with $w=1\mathrm{\:\mu m}$ and $l=20\mathrm{\:nm}$. If the weak link behaves as a tunnel barrier with low transparency across its entire area,  
we expect the Ambegaokar-Baratoff relation, $I_cR_N=\pi\Delta/2e$~\cite{AmbBar_1963}, to hold near zero temperature, where $R_N$ is the normal state resistance of the weak link.  Instead, we find $I_cR_N\approx 0.3\cdot\pi\Delta/2e$ with $\Delta_{\mathrm{NbN}}=2.03\mathrm{\: meV}$ inferred
from the critical temperature of the film [Fig.~\ref{fig1}] and $R_N$ is inferred from the slope of $I-V$ curves above $I_c$ [Fig.~\ref{fig2}].

The suppression of the $I_cR_N$ product may indicate that electrons dwell in the weak link over a time $\sim \hbar /E_{\rm Th}$ exceeding the scale $\sim \hbar /\Delta$ defined by the superconducting gap. In the case of electron diffusive motion across the weak link, the conventionally defined Thouless energy is $E_{\rm Th}=D/l^2$, where the diffusion coefficient $D=v_Fl_e/3$ is proportional to the mean free path $l_e$ and the velocity of the electron $v_F$ at the Fermi level $v_F\approx (0.7-2)\times 10^6{\rm m/s}$ in NbN~\cite{Kern_2024}). A theory applicable to ``long'' weak links with $E_{\rm Th}<\Delta$ was developed in Ref.~\cite{dubos2001_2} and recently confirmed and extended in Ref.~\cite{ostrovsky2025}. An experiment~\cite{dubos2001_2} performed with well-characterized long metallic links in the regime $E_{\rm Th}\lesssim 0.1\Delta$ agreed well with the theory. We can check whether this theory is consistent with our observations by applying it to extract a mean free path from the data. This results in $l_e=0.03-0.5~{\rm nm}$, comparable to NbN inter-atomic distance. The extremely short mean free path rules out conventional electron diffusion and indicates electron localization within the weak link. The theory of the Josephson effect in this regime is far less universal than those for the two opposite limits of tunneling across a uniform barrier~\cite{AmbBar_1963} and diffusive electron motion~\cite{dubos2001_2,ostrovsky2025}, respectively. In Section IV, we will discuss candidate models consistent with the entirety of our data.

Extended data in the Supplemental Information [Fig.~S4] shows the measured critical currents of superconducting weak links as a function $w$ and $l$. We find a tunable range from $I_c\sim 20\mathrm{\:n A}$ to $\sim 100\mathrm{\:\mu A}$. For sufficiently long weak links, $l\sim 50\:\mathrm{nm}$, we did not detect a critical current within the sensitivity of our measurements [Fig.~\ref{fig2}e and ~\ref{fig2}f]. The differential resistance $dV/dI$ at low biases, $V<V_c$, exceeded 100 M$\Omega $, with the characteristic bias substantially exceeding the gap of the leads, $V_c\gg2\Delta_{NbN}/e$, see figs.~\ref{fig2}e and ~\ref{fig2}f. We characterize this junction behavior as insulating one. It is currently not well understood why weak links display insulating behavior when the barrier is sufficiently long.\\
\indent To realize a transmon qubit (Section III) we will use weak link parameters that results in superconducting behavior [similar to Fig.~\ref{fig2}c].

\section{III. Planaron qubit }
Next, we report measurement of a transmon qubit fabricated from a single film of NbN, dubbed the '$planaron$'. The device is fabricated in the same way as detailed in Sec.~II, except the DC leads are replaced by capacitor pads to realize a transmon geometry. We implement a weak link of dimensions $w\sim 1\:\mathrm{\mu m}$ and $l\sim 30\:\mathrm{nm}$, see Fig.~\ref{fig3}a, which results in a device with a measured anharmonicity of $\sim 235 \mathrm{\ MHz}$. In the Supplemental Information we show that different weak link geometries can lead to orders of magnitude lower anharmonicity, despite having a similar inductance and shunting capacitance. We use this to create highly-lumped resonators with anharmonicity of $\sim 10$ Hz [Fig.~S7]. The large difference in anharmonicity, tuned by junction geometry, is currently not well understood and is subject of future work.

To probe the planaron qubit, we directly couple it to a transmission line in a hanger geometry without a readout resonator [Fig.~\ref{fig3}a]. Omitting the readout resonator makes it easier to characterize devices with unknown frequency and anharmonicity. Spectroscopy is instead performed by probing the complex transmission amplitude, $S_{21}$, of a signal propagating along the transmission line. 
First, we demonstrate that the device operates as a qubit with anharmonicity $\alpha$ larger than the decoherence rate $\kappa_t$. This can be verified using only single-tone spectroscopy: when $\alpha \gg \kappa_t$, the $|0\rangle \rightarrow |1\rangle$ transition do not overlap with the $|1\rangle \rightarrow |2\rangle$ transition, and behaves as a two-level system. The drive cannot populate the mode with more than one photon and the system therefore becomes saturated when the drive strength $\Omega_R$ is similar to the decay rate $\kappa_t$. This saturation results in the lineshape becoming continuously "squashed" [Fig.~\ref{fig3}b] in the complex plane as the drive power is increased \cite{chang2007,Astafiev2010,winkel2020}:

\begin{align}
\label{squash}
\delta S_{21}=\frac{\kappa_c}{\kappa_t}\frac{1+2i\Delta/\kappa_t}{1+(2\Delta/\kappa_t)^2+2(\Omega_R/\kappa_t)^2}.
\end{align}

\noindent Here $\delta S_{21}$ is the change in the transmitted signal due to the resonance, $\Delta$ is the detuning between the probe frequency and the transition, and $\kappa_c$ is the coupling to the transmission line. For sufficiently strong drives, the response of the two-level system becomes negligible, and the change in transmission, $\delta S_{21}$, approaches zero. Unlike the weakly anharmonic oscillator, there is no shift in the resonance frequency and no discontinuity in the frequency-dependent transmission at high power. As shown in Fig.~\ref{fig3}b and ~\ref{fig3}c, we find good agreement with the expected response of a system with $\alpha \gg \kappa_t$, Eq.~(\ref{squash}). 

We extract the total linewidth of the qubit to be $\kappa_t/2\pi \approx  15\:\mathrm{MHz}$. The underlying decoherence mechanisms are not yet fully understood. We leave a detailed investigation of intrinsic junction loss, spurious mode coupling, and design optimization to future studies.
The total decoherence rate, however, is much larger than the coupling to the transmission line $\kappa_c/2\pi \approx 75 \:\mathrm{kHz}$, which presents a challenge to our measurements. When $\kappa_t \gg \kappa_c$, the change in transmission on resonance is small, about $5\times10^{-3}$ for this device. To make the resonance clearly visible, the background transmission is subtracted. The background is found by probing at high power ($\Omega_R \gg \kappa$) where the device does not affect the transmitted signal.

Next, we perform two-tone spectroscopy to determine the anharmonicity of the planaron qubit. The $|1\rangle$ state is populated by continuously driving at $f_d =f_{01}$. Simultaneously, the frequency of a second spectroscopy tone is swept to identify $f_{12}$. To clearly identify the strongly undercoupled $|1\rangle \rightarrow|2\rangle$ transition, the response of the system with the $|0\rangle \rightarrow|1\rangle$ drive off is subtracted. As shown in Figs.~\ref{fig3}d and \ref{fig3}e, a second peak appears $235$ MHz below the $|0\rangle \rightarrow|1\rangle$ transition, which is attributed to the presence of the second excited state. As the population of the $|1\rangle$ state increases with higher drive power, the visibility of the $|1\rangle \rightarrow|2\rangle$ transition also increases. At the highest drive powers, however, the $|1\rangle \rightarrow|2\rangle$ transition disappears. We attribute this to significant population of highly excited states due to the presence of the strong drive \cite{sank2016, dumas2024}.

In the presence of a strong resonant drive, we also observe the $|0\rangle \rightarrow|1\rangle$ and $|1\rangle \rightarrow|2\rangle$ transition each split into a pair of doublets with their splitting proportional to the drive amplitude $\Omega_R \propto \sqrt{P_{d}}$. This is the well-known Aulter-Townes effect, where the drive splits each bare state into a pair of dressed states \cite{autler1955}. These transitions can be measured spectroscopically, shown in Fig.~\ref{fig3}d as a function of drive power. The predicted transition frequencies (dashed lines) match the theoretical model well, $\delta f= f_q\pm \sqrt{4\kappa P_d/hf_{01}}$ \cite{Astafiev2010,winkel2020}. Fitting to the sidebands further allows us to calibrate total attenuation of the setup, resulting in $-135\:\mathrm{dB}$, consistent with what is expected from lumped and distributed attenuation in the input lines. We note that only the lower branch of $|1\rangle\rightarrow|2\rangle$ transition can be properly fitted due to the interference with $|0\rangle\rightarrow|1\rangle$ splitting at high drive powers.

Finally, based on the designed charging energy of $E_C/\hbar = 2\pi\times 293\mathrm{\ MHz}$ and measured frequency $f_q=7.945\:\mathrm{GHz}$, we extract an estimated Josephson energy of $E_J/\hbar =2\pi \times 26.15\:\mathrm{GHz}$ \cite{Koch2007-pu}. This places the planaron in the transmon regime with $E_J/E_C\sim90$. The anharmonicity, $\alpha \simeq - E_C$, is thus predicted to be $306 \mathrm{\ MHz}$ where additional $\sim4$\% comes from numerical diagonalization \cite{Koch2007-pu}. The measured anharmonicity of $235 \mathrm{\ MHz}$ is modestly reduced compared to the prediction, which assumes a low-transparency S-I-S junction with same inductance. This deviation cannot be explained by stray inductance only; we estimate that the inductance of our leads reduces $\alpha/2\pi$ by only $\sim 12 $ MHz \cite{willsch2024} (see Supplemental Information). The discrepancy is therefore more likely due to the microscopic nature of the NbN weak link. Transparent transmission channels, for example, are known to reduce anharmonicity by up to a factor of four \cite{kringhoj2018}.

\section{IV. Evidence for highly transparent channels in planar weak links}
In this section, we further investigate our NbN weak links in an effort to understand their microscopic nature. In addition to the $I_cR_N$ products at the lowest temperature, we measured the evolution of the critical current with temperature, $I_c(T)$, and the DC response of a weak link to microwave irradiation. We will see that the measurement results prompt us to conjecture that microscopically the weak link material is close to the disorder-driven localization transition.

As mentioned in Sec.~II, the electron diffusion theory indicates that the mean free path $l_e$ is of the order of NbN inter-atomic distance. That points towards a small value of the product $k_Fl_e\lesssim 1$ and therefore, to the insulating side of the Anderson localization transition. 

Contrary to band insulators, the insulating state of a strongly disordered conductor does not feature a gap in the electron density of states. Consequently, tunneling through a thin layer of an Anderson insulator is enhanced by the presence of resonant levels inside the layer with energy close to the Fermi level in the leads. Optimal configurations of the resonant levels can provide highly transparent channels that dominate the conduction across a disordered insulator \cite{lifshitz1979}. Such configurations include single levels for thin layers, while more complicated arrangements dominate conduction of thicker ones~\cite{Beasley_1987,LGRS_1988,larkin1988}. 
\begin{figure}[t]
	\includegraphics[width= 3.5 in]{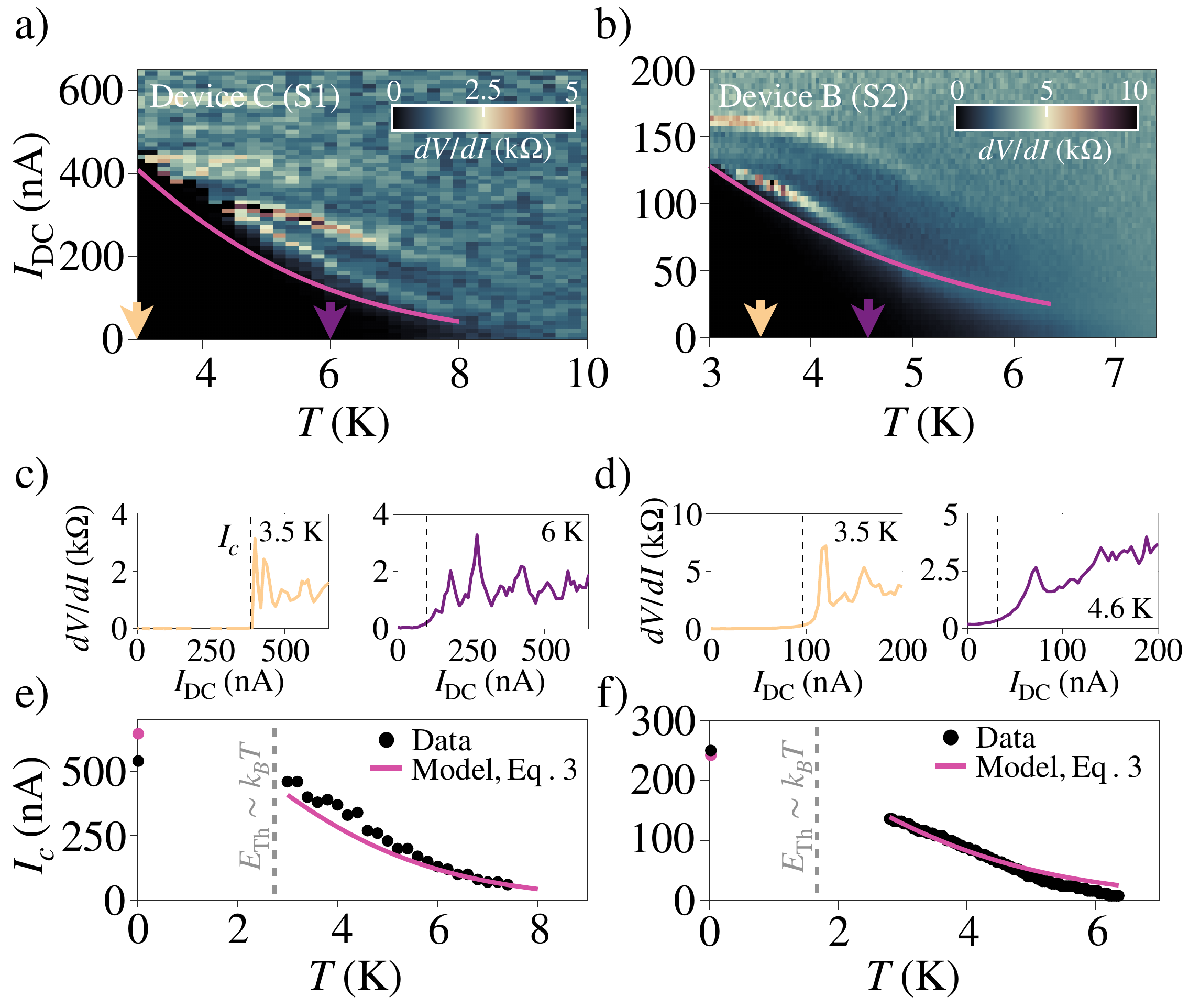}
	\caption{\textbf{Critical current evolution with temperature }
 a, b) A monotonic decrease in critical current, $I_c$, is observed when the temperature is increased from 3 K to 10 K. We define $I_c$ as the boundary between a measured zero differential resistance, $dV/dI$, and the onset of measurable finite $dV/dI$. Both weak links inherit large-gap properties of NbN, resulting in a measured critical current that goes to zero near $\sim 7-8$ K. c, d) Line cuts of panels a, b) are taken along the directions indicated by orange and purple arrows; they illustrate how critical current at each temperature, $I_c(T)$, is extracted  from differential resistance measurements. Vertical dashed lines in panels c, d) mark the threshold for $I_c$. We note the transition becomes progressively less sharp at elevated temperatures. In this regime, we define the threshold as approximately the onset of a \textit{change} in differential resistance, i.e. when $d^2V/dI^2\neq 0$. e, f) Critical current evolution with temperature is modeled assuming we are in the diffusive limit using Eq.~(\ref{critical current}) with the diffusion coefficient, $D$, as the only one free parameter (pink curve). The best fit to this data results in small values of e) $D$: $1.1 \:\mathrm{cm^2/s}$ and f) $0.2 \:\mathrm{cm^2/s}$, see main text for discussion on effects of localization. Close to zero temperature we plot the separately measured values of $I_c$ (black solid circles) alongside the predicted values (pink solid circles), assuming diffusive the limit.
			}
	\label{fig4}
\end{figure}

Addressing the Josephson effect in tunneling across a disordered insulator in the limit of strong localization, we consider the simplest case of single-level configurations. These configurations are dominant in the case of short, strongly disordered links. To find the average over configurations, we combine the results known for the Josephson current through a single resonant level~\cite{Beenakker_1992} with the proper averaging procedure developed for the normal state~\cite{LGRS_1988,larkin1988}. The low-temperature ($k_BT< E_{\rm Th}$) result can be cast in the form

\begin{equation}
I_J(\varphi)R_N=\frac{E_{\rm Th}}{e}\left\{\ln\frac{4\Delta}{E_{\rm Th}}+\frac{1}{\pi}\left[\sin^{-1}\!\!\left(\sin\frac{\varphi}{2}\right)\right]^2\right\}\sin\varphi.
\label{Eq:restunn}
\end{equation}

Here $1/E_{\rm Th}$ is the electron dwelling time in a resonant state localized in the middle of the insulating layer. Defined this way, $E_{\rm Th}$ equals to the level width determined by an electron escape into normal leads, in agreement with the general definition of the Thouless energy. In the derivation, we assumed $E_{\rm Th}\ll\Delta$, in line with the experimental data for the $I_cR_N$ product. We note that despite the small $I_cR_N$ product, Eq.~(\ref{Eq:restunn}) has a strong non-sinusoidal component. Qualitatively, it arises from the presence of highly-transparent transport channels with energy width $\sim E_{\rm Th}$, allowing for phase-sensitive mid-gap Andreev levels. The deviations of $I_cR_N$ from Ambegaokar-Baratoff theory diminish with temperature, and it becomes applicable at $T$ closer to $T_c$, once $\Delta(T)\lesssim 2k_BT$; close to $T_c$, the theoretically expected derivative $d[I_c(T)R_N]/dT\approx -635\:\mu{\rm eV/K}$ \cite{tinkham2004}.

We measured $dV/dI$ for two representative weak links as a function of temperature from 3 K to 10 K. Figure~4 shows the extracted critical current approaches zero near $T\sim 7-8\:\mathrm{K}$. This indicates the weak links inherit the large-gap of NbN. As suggested by the previously calculated $I_cR_N$ products, the temperature dependence of the critical current does not exhibit the typical BCS superconductor-insulator-superconductor (S-I-S) junction behavior. Comparing the experimentally found relation $I_cR_N\approx 0.3\cdot\pi\Delta/2e$ (see Section II) with Eq.~(\ref{Eq:restunn}), we estimate $E_\mathrm{{Th}}/\Delta(0)\approx 0.14$, in agreement with the expectation of a small value of $E_\mathrm{{Th}}$. However, the minimal model of a disordered insulator runs into a problem explaining the $I_c(T)$ dependence at $T\to T_c$: the theoretically expected value of $|dI_c(T)/dT|\approx 500\:{\rm nA/K}$ is over 3 times higher than the highest measured values of $|dI_c(T)/dT|\approx 150\:{\rm nA/K}$, see Figs. \ref{fig4}a, \ref{fig4}b (in the theoretical estimate, we used $R_N\sim 1.3\:{\rm k\Omega}$, see Fig. \ref{fig2}d). 
\newline

\begin{figure*}[t]
	\includegraphics[width= 7.1 in]{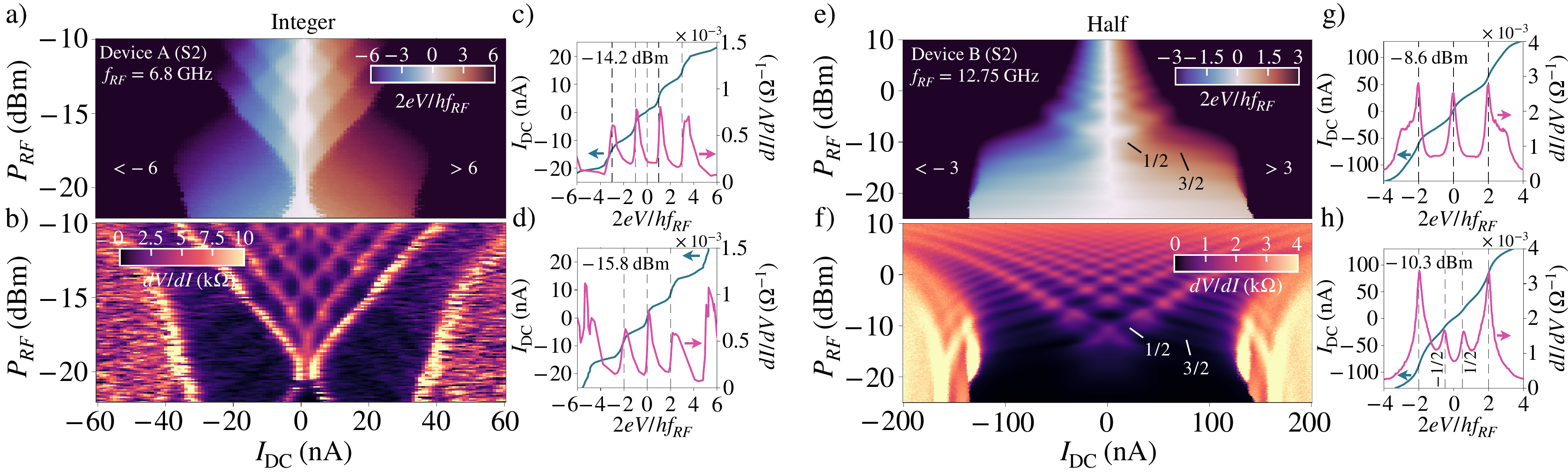}
	\caption{ \textbf{Integer and half Shapiro features} a, b) DC response under the influence of RF radiation fixed at $6.8\: \mathrm{GHz}$. The irradiation leads to integer steps in the measured voltage in units of the microwave field, $2eV/hf_{RF}$, with corresponding lobes of zero differential resistance, $dV/dI$, between steps. c, d) Line cuts taken at two intermediate microwave powers show rounded integer Shapiro steps in $I-V$ and corresponding peaks in measured conductance, $dI/dV$. e, f) Maps of measured voltage in units of microwave field and measured differential resistance show rounded half Shapiro steps develop in a representative sample with a fixed frequency of $12.75 \:\mathrm{GHz}$. g, h) Line cuts at two intermediate microwave powers show both integer and half steps with corresponding conductance peaks.
    } 
	\label{fig5}
\end{figure*}

The above difficulty in explaining the observed temperature dependence of $I_c$ prompts us to revisit the model of a coherent diffusive weak link. If one dispenses with the notion of localization, it allows one to explain the overall $I_c(T)$ behavior.
The supercurrent in a diffusive superconductor-normal-superconductor (S-N-S) junction can be described by superconducting correlations which bleed into the normal barrier, mediated by coherent Andreev reflection at the N-S interface. At temperatures larger than the Thouless energy, the critical current in the diffusive limit is exponentially suppressed by the factor $I_c\sim \exp(-l/l_T)$, where $l_T = \sqrt{\hbar D/2\pi k_B T}$ is the characteristic thermal length, $l$ is the junction length, and $D$ is the diffusion constant. This exponential suppression helps to explain the low values of $|dI_c(T)/dT|$ when $T$ is in the vicinity of $T_c$. The full theoretical prediction for the dependence $I_c(T)$ is described by~\cite{dubos2001}
\begin{equation}
\label{critical current}
I_cR_\mathrm{N}=\frac{64\pi k_B T}{e}\sum^\infty_{n=0} \frac{l/l_{\omega_n}\Delta(T)^2 \exp(-l/l_{\omega_n})}{(\omega_n +\Omega_n+\sqrt{2(\Omega^2+\omega_n\Omega_n)})^2}. 
\end{equation}
Here we have defined $\Omega_n=\sqrt{\Delta^2+\omega_n^2}$ and $l_{\omega_n}=\sqrt{\hbar D/2\omega_n}$ where $\omega_n$ is the $n^{\mathrm{th}}$ Matsubara frequency given by $\omega_n=(2n+1)\pi k_B T$.

We numerically solve Eq.~(\ref{critical current}) to obtain $I_c(T)$ with input parameters from our experimental data. We use $R_N$ values extracted directly from $I\mathrm{-}V$ curves and use the junction length $l$ measured by atomic force microscopy. We assume the superconducting gap is well-described by strong-coupling limit BCS theory; $\Delta_0=1.96 k_B T_c$ ~\cite{sim2017} and $\Delta(T)=\Delta_0\sqrt{(1-T^2/T_c^2)}$.  We take the critical temperature to be $T_c=12 \:\mathrm{K}$ based on the measurement of the films in Section I. This leaves the diffusion coefficient as the only free parameter, which we extract by fitting $I_c(T)$ for both weak links. Equation~(\ref{critical current}) is only applicable in the limit $k_BT>E_{\mathrm{Th}}$, so we must use the calculation from Ref. \cite{dubos2001} to include the zero-temperature critical current in the fit. We compare this result to the measured $I_c$ at base temperature and find good agreement with the model [Fig.~\ref{fig4}a and ~\ref{fig4}b]. 

We caution that while the diffusion model can predict the temperature dependence of $I_c$ reasonably well, it disregards the localization effect. This is not fully justified because of the very low extracted diffusion coefficient ($D=0.2-1.1\mathrm{\:cm^2}$) and extremely short estimated electron mean free path $l_e=0.03-0.5~{\rm nm}$; this suggests localization in the link. 

The finite-temperature suppression of the critical current, $I_c\sim \exp(-l/l_T)$, comes from thermal averaging of the electron propagation amplitudes, which have energy-dependent phases. In the opposite limit of strong localization and prevalence of single-level configurations such phase averaging effect is suppressed -- which ultimately creates a difficulty with explaining the observed $I_c(T)$ behavior. A possible resolution of the conflict presented by the two models is that in reality our system is close to the localization transition, so between the applicability domains of the two models.
\newline
\indent Next, we probe the DC response of the weak link under the influence of a radio frequency (RF) drive. The drive modifies the $I-V$ characteristic, resulting in Shapiro steps which can be used to gain insight into the current-phase relation of the weak link.
Figure 5 summarizes the results of two representative devices with extended data presented in Supplemental Information. We explore the measured DC voltage [Fig.~\ref{fig5}a] with differential resistance maps [Fig.~\ref{fig5}b] as a function of applied RF power, $P_{RF}$, and DC bias current for a fixed RF frequency, $f_{\mathrm{RF}}$. We observe rounded steps resembling conventional integer Shapiro steps for $V=mhf_{RF}/2e$ with $m=1,2,3,...$, as expected for a weak link with a sinusoidal current-phase relation. Fig.~\ref{fig5}c and ~\ref{fig5}d display line cuts taken at two different RF powers. Notably, as shown in Fig.~\ref{fig5}e and ~\ref{fig5}f, we also observe Shapiro-like features which occur at fractions of the aforementioned voltages when this device is irradiated at higher frequency, 12.75 GHz. These fractional Shapiro steps are indicative of harmonics in the current-phase relation which may arise from the presence of transparent transmission channels. In the diffusive limit, transmission is high at any energy within the gap, resulting in a non-sinusoidal current-phase relation~\cite{houzet_2008}. One may see indeed that in the strong localization limit the current-phase dependence $I(\varphi)$ in Eq.~(\ref{Eq:restunn}) contains higher harmonics of phase, which would also lead to the fractional Shapiro steps. Therefore this observation cannot distinguish between the diffusive limit and the strong localization limit.
Additional data in Supplemental Information show that half and integer steps are present in both devices. 

We note that half-steps may also be present in junctions where stray inductance is significant \cite{likharev1986}. The leads will provide a small but finite inductance primarily from the sheet inductance, which is measured to be $\sim 10-20\:\mathrm{pH/\square}$ [Fig.~\ref{fig1}d], resulting in a total stray inductance of $100-200 \:\mathrm{pH}$. This is small compared to the estimated weak link inductance of $\sim 5-10 \:\mathrm{nH}$, so we conclude that stray effects are likely not responsible for the observed half-integer rounded Shapiro step.

The presented transport data is consistent with the assumption that the parameters of our NbN links were close to the localization transition: on the one hand, a small $I_cR_N$ product and the form of the $I_c(T)$ dependence can be explained by electron diffusion. On the other the extracted electron mean free path $l_e$ is short on the scale of the Fermi wavelength, suggesting localization. The microwave response of the weak links demonstrates the presence of channels transparent for electron propagation, despite the short electron mean free path.

\section{V. Conclusion and outlook}
In conclusion, this article presents a novel approach for producing weak links and associated superconducting qubits. We have shown that by leveraging the thickness-tunable superconductor-insulator transition in NbN, we can simplify fabrication processes using an all-subtractive ALD/ALE method and realize planar weak links without the need for oxide layers or dielectric barriers. We utilize this method to produce a transmon qubit with moderately reduced anharmonicity. The reduction possibly stems from the presence of highly transparent channels in the link. Additional evidence for high transparency channels is the observation of half-integer Shapiro features. We note that the presence of such channels is possible for disordered short links even in the strong localization limit.
\newline
\indent In a marked difference from the coplanar waveguide resonators, the transmon we studied here have relatively large dissipation, which is currently not understood. Indeed, CPW resonators fabricated from bulk films and subsequently thinned by ALE, reveal a high internal quality factor of $\sim 10^6$ . We leave a detailed investigation of the nature of potential junction loss to future studies. 
\newline
\indent Our weak links may be applied to a wide range of superconducting circuits. Using materials with larger gap than aluminum may be particularly suited for the millimeter wave regime and temperatures above 1~K. This investigation also invites further exploration of fundamental questions relating to the SIT in finite-sized samples. \\

\begin{acknowledgments}
\textit{Acknowledgments:} The authors thank Nicholas R. Poniatowski, Charles M. Marcus, and Aharon Kapitulnik for helpful discussions and feedback on the manuscript. We thank Yufeng Wu and Chaofan Wang for  discussions on fabrication processes and NbN film characterization. We thank Patrick Winkel for discussions on resonance fluorescence of a transmon. We thank Yong Sun for assistance with developing ALE recipes.
This research was sponsored by  DARPA under grant no. HR0011-24-2-0346, by the Army Research Office (ARO) under grants no. W911NF-22-1-0053 and W911NF-23-1-0051, by the U.S. Department of Energy (DoE), Office of Science, National Quantum Information Science Research Centers, Co-design Center for Quantum Advantage (C2QA) under contract number DE-SC0012704,  by the Air Force Office of Scientific Research (AFOSR) under award number FA9550-21-1-0209, and by the Office of Naval Research (ONR) under award number
N00014-22-1-2764 and N00014-23-1-2121. We acknowledge support from research grants (Projects No. 43951 and No. 53097) from VILLUM FONDEN. The views and conclusions contained in this document are those of the authors and should not be interpreted as representing the official policies, either expressed or implied, of the DARPA,  ARO, DoE, AFOSR or the US Government. The US Government is authorized to reproduce and distribute reprints for Government purposes notwithstanding any copyright notation herein. Fabrication facilities use was supported by the Yale Institute for Nanoscience and Quantum Engineering (YINQE) and the Yale Univeristy Cleanroom. 
\end{acknowledgments}
\section{Author contributions}
C.B. and M.D. conceived the research and designed the experiments. C.B. deposited the ALD NbN films with assistance from D.W. P.K. prepared and annealed the sapphire wafers, and C.B. and E.{\"O}. fabricated the DC and RF devices. C.B., E.{\"O}., and T.C. performed microwave simulations, microwave measurements and analyzed the data. DC measurements were carried out by C.B., E.{\"O}., J.Z., C.K., and D.W., with contributions from S.V. and H.T. The manuscript was written by C.B., E.{\"O}., T.C., L.G., and M.D., with input from all authors. 
\section{Competing interests}
The authors declare no competing interests. 
\section{Data availability}
The data supporting the findings of this study are available upon request.

\bibliographystyle{IEEEtran}
\bibliography{refs}

\newpage

\appendix

\input{supplement}

\end{document}

%% file: supplement.tex
\preprint{APS/123-QED}

\title{Supplemental information for ``A transmon qubit realized by exploiting the superconductor-insulator transition"}

\author{C.~G.~L.~B\o{}ttcher}
\thanks{Email: charlotte.boettcher@stanford.edu, and \\
michel.devoret@yale.edu.}
\affiliation{Department of Applied Physics, Yale University, New Haven, CT 06520, USA}
\affiliation{Department of Applied Physics, Stanford University, Stanford, CA 94305, USA}
\author{E.~{\"O}nder}
\affiliation{Department of Applied Physics, Yale University, New Haven, CT 06520, USA}
\author{T.~Connolly}
\affiliation{Department of Applied Physics, Yale University, New Haven, CT 06520, USA}
\author{J.~Zhao}
\affiliation{Center for Quantum Devices, Niels Bohr Institute, University of Copenhagen, 2100 Copenhagen, Denmark}
\author{C.~Kvande}
\affiliation{Center for Quantum Devices, Niels Bohr Institute, University of Copenhagen, 2100 Copenhagen, Denmark}
\affiliation{Department of Physics, University of Washington, Seattle, Washington 98195, USA}
\author{D.~Q.~Wang}
\affiliation{Department of Electrical Engineering, Yale University, New Haven, CT 06520, USA}
\author{P.~D.~Kurilovich}
\affiliation{Department of Physics, Yale University, New Haven, CT 06520, USA}
\affiliation{Department of Applied Physics, Yale University, New Haven, CT 06520, USA}
\author{S.~Vaitiek\.{e}nas}
\affiliation{Center for Quantum Devices, Niels Bohr Institute, University of Copenhagen, 2100 Copenhagen, Denmark}
\author{L.~I.~Glazman}
\affiliation{Department of Physics, Yale University, New Haven, CT 06520, USA}
\affiliation{Yale Quantum Institute, Yale University, New Haven, Connecticut 06511, USA}
\author{H.~X.~Tang}
\affiliation{Department of Electrical Engineering, Yale University, New Haven, CT 06520, USA}
\author{M.~H.~Devoret}
\thanks{Email: charlotte.boettcher@stanford.edu, and \\
michel.devoret@yale.edu.}
\affiliation{Department of Applied Physics, Yale University, New Haven, CT 06520, USA}
\affiliation{Department of Physics, UCSB, Santa Barbara, CA 93106, USA}
\affiliation{Google Quantum AI, 301 Mentor Dr, Goleta, California 93111, USA}

\date{\today}

\renewcommand{\thefigure}{S\arabic{figure}}

\section{A. Atomic layer deposition and etching of niobium nitride}
We grow niobium nitride (NbN) films using atomic layer deposition (ALD) in an Ultratech/Cambridge Fiji G2 plasma system. During each cycle tris(diethylamido)(tert-butylimido)niobium(V) (TBTDEN) precursor is introduced into the chamber. Shortly after, $\rm H_2/\mathrm{N}_{2}$ plasma is ignited in the chamber and reaction is facilitated by a high chamber and chuck temperature, in our case 400 \textdegree{C}. We grow our films on a 2-inch annealed HEMEX sapphire wafer with a thickness of $650 \mathrm{\ \mu m}$. Wafer preparation and anneal recipe is detailed in Ref.  \cite{ganjam2024}. The final NbN thickness is typically $20$ nm, unless otherwise noted. 
\newline
\begin{figure*}[t]
	\includegraphics[width= 7 in]{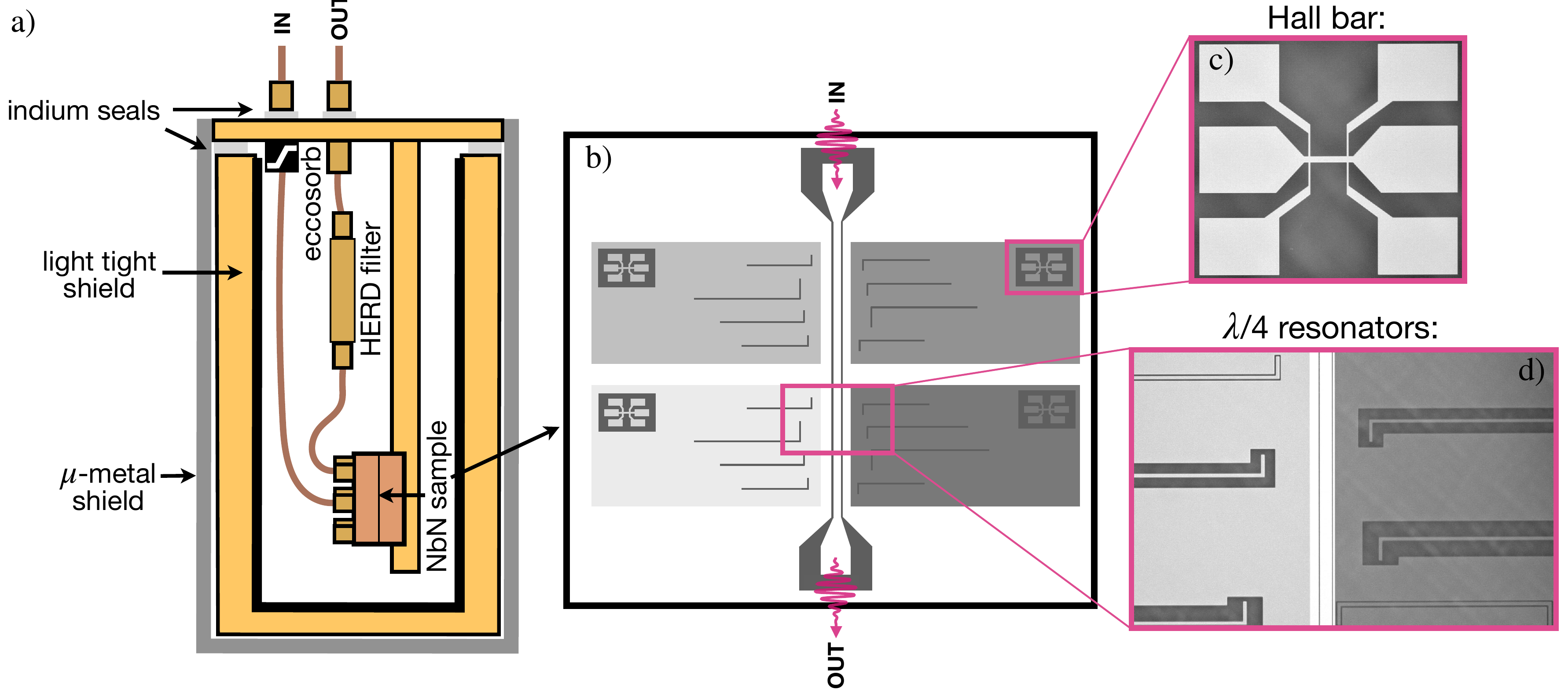}
	\caption{\textbf{Microwave setup and hybrid RF, DC sample layout} a) Measurement setup of our NbN samples inside a dilution refrigerator with light tight and $\mu \mathrm{-metal}$ shields. b) A sketch of our multi-patch chip (not to scale). c) Optical image of a Hall bar. d) Optical image of microwave resonators coupled to a transmission line. 
			}
	\label{figS1}
\end{figure*}
We then use atomic layer etching (ALE) to etch the deposited NbN films with sub-nanometer precision. The ALE process is carried out in a reactive ion etching inductively coupled plasma (RIE–ICP) system. In the first cycle we introduce $\mathrm{SF}_6$ diluted with $\mathrm{H_2}$ into the chamber. Subsequently, a low-power plasma is applied at 30 W for 10 seconds, followed by a final step where the system is pumped for byproducts before the next cycle. This process facilitates controlled etching without the use of ICP. The main chamber is maintained at 15\textdegree{C}. Each cycle removes, on average, less than one nanometer of material. This technique allows precise tuning of the local thickness of the NbN film and is used in the fabrication of the DC and RF devices reported in this work.

Unless otherwise noted, all samples undergo a global ALE treatment that removes approximately 3 nm of NbN and oxide combined. This step improves surface roughness and results in minimal microwave losses (appendix B.). For selective etching into specific device geometries, lithography steps are introduced. To fabricate our devices, we define patterns using electron-beam lithography. The patterned areas are then etched to form devices. Compared to traditional fabrication methods that rely on lift-off processes, this approach is based entirely on film subtraction, potentially allowing cleaner fabrication and processing. Finally, because all samples are NbN only, they are compatible with aggressive cleaning methods, such as Buffered Oxide Etching (BOE). This is used as a final step to remove residual carbon and fluoride contaminants \cite{anferov2024_2}.

\begin{figure}[t]
	\includegraphics[width= 2.5 in]{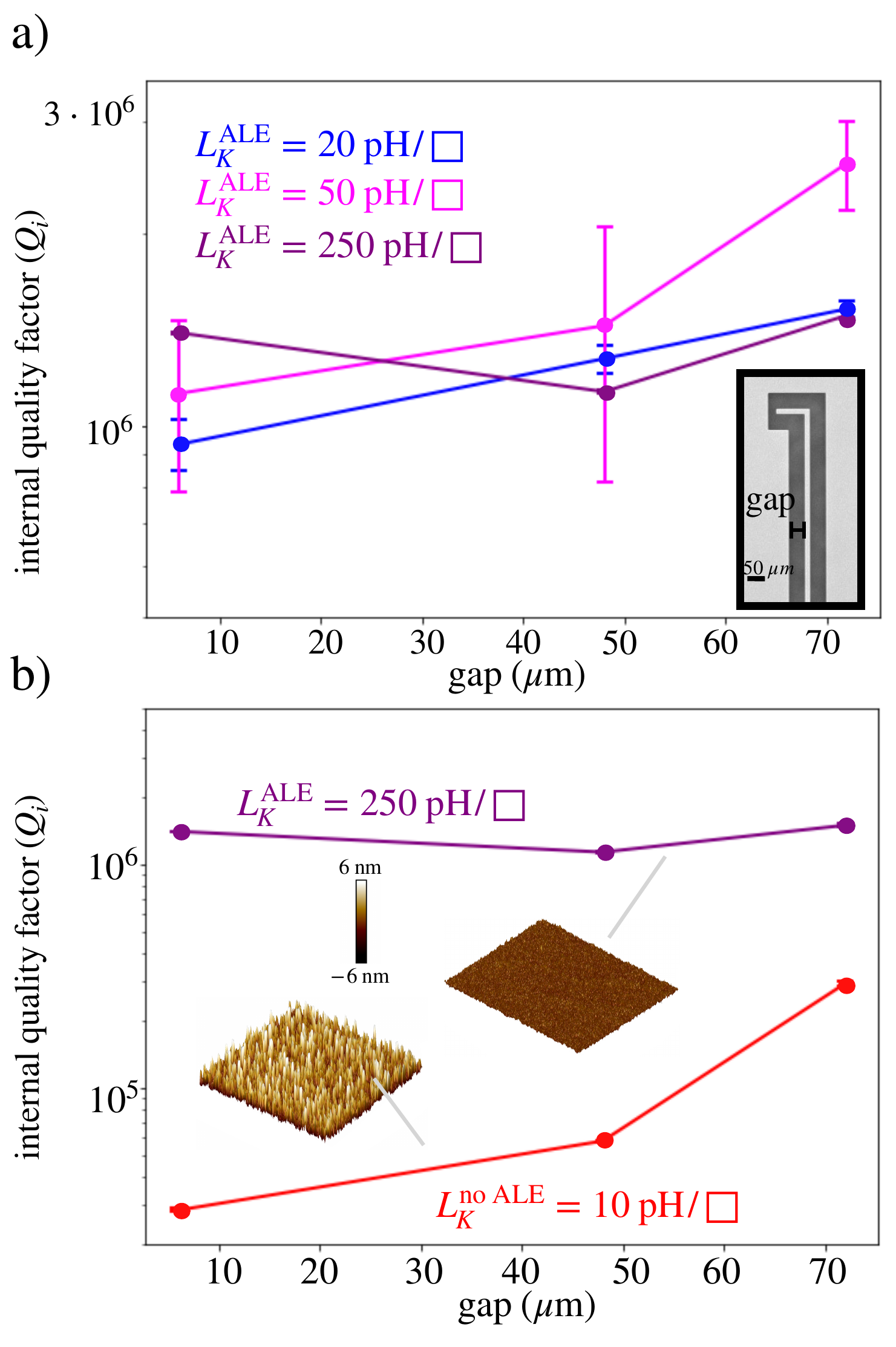}
	\caption{\textbf{Measured internal quality factor across devices} a) Internal quality factors extracted from resonators fabricated on patches of different thickness are plotted as a function of resonator gap (inset : optical image of a CPW resonator). Each color signifies a different patch of different thickness and thus different kinetic inductance. The labeled kinetic inductance values are extracted from measured frequencies of the CPW resonators and compared to finite element simulations. b) Internal quality factors for resonators fabricated on two different patches are plotted as a function of resonator gap. The purple line represents data from resonators fabricated on a patch of film whose kinetic inductance is $250 \: \mathrm{pH}$ and has been treated with ALE. The red line represents data from resonators fabricated on a patch of film whose kinetic inductance is $10 \: \mathrm{pH}$ and has not been treated with ALE. AFM images with a scale bar for reference is presented as an inset with the respective AFM image pointing towards the relevant data set.
			}
	\label{figS2}
\end{figure} 

\section{B. Transport and microwave characterization of niobium nitride}
To evaluate ALD-grown NbN films and effects of ALE, we performed a comprehensive set of characterizations, including measurements of surface morphology, transport properties, and microwave measurements. 

\subsection{Measurement Setup}
Our standard chip size is 7 mm x 7 mm cut from our 2-inch wafers. The chip is mounted in a copper package with an air gap below and above the chip. The package is mounted on gold plated copper brackets for good thermal contact to the mixing chamber of our dilution refrigerator. The sample box is shielded from thermal photons by a light tight shield \cite{connolly2024} closed tightly with indium seals and a $\mu$-metal shield. Our input line uses Eccosorb filters in series with attenuators that provide broadband filtering to prevent high-frequency thermal photons from reaching the device. We use high-energy radiation dissipating (HERD) filters to cleanly transmit desired frequencies from our device while cutting out unwanted ones in our output line [Fig. \ref{figS1}a]. We note that the transmon qubit is measured using the same setup.
\newline
\indent The Hall bars and microwave CPW resonators are fabricated by etching multiple patches of different thicknesses on a single chip [Fig. \ref{figS1}], enabling multiplexed readout of resonators with varied parameters. There are four different patches on every 7 x 7 mm chip. Each patch is made by an e-beam lithography step followed by ALE to etch the exposed area on the chip. This process is repeated for each patch, resulting in four patches of varied thickness and hence kinetic inductance. Kinetic inductance ranges from $10 \:\mathrm{pH/\square}$ to $250 \:\mathrm{pH/\square}$ within a single chip. For the Hall bar and CPW structures, another lithographic step is used followed by a long ALE step to remove the entire NbN film in CPW gaps and to isolated devices [Fig. \ref{figS1}].

\subsection{Hall bars}
To characterize transport properties of our films, we use Hall bar structures that enable measurement of resistance per square values, $R_s$. Hall bars are patterned alongside the resonators on every patch and etched simultaneously with the resonators. The Hall bar width has same dimensions as CPW signal trace, which allows us to include relevant finite-size effects in our characterization. The measured normal state resistance per square, $R_N$, is correlated with the kinetic inductance per square, $L_K$, of the material through the Mattis-Bardeen relation:
\begin{align}
    L_K= \frac{\hbar R_N}{\pi \Delta(0)}  
\end{align}
We use $\Delta(0) = 1.96 \, k_B T_c,$~\cite{sim2017}, zero-temperature strong-coupling limit BCS gap where $k_B$ is the Boltzmann constant and $T_c$ is the critical superconducting transition temperature. We perform standard low-frequency lock-in measurements to measure the superconducting transition and $R_N$ for each film of a particular thickness.

\subsection{Microwave CPW Resonators}
We fabricate between three to four coplanar waveguide (CPW) resonators on each patch, with target frequencies in a wide band between $5-9$ GHz. Within a patch, we design the resonators to be $\sim 200$ MHz apart. Each resonator in a patch has a different gap from the signal trace to the ground plane. We use 6$\mu \mathrm{m}$, 24 $\mu \mathrm{m}$, 48 $\mu \mathrm{m}$ and 72 $\mu \mathrm{m}$. An additional resonator with a 6 $\mu \mathrm{m}$ gap is included and intentionally over-coupled to facilitate the identification of resonance dips during the transmission measurements over a wide frequency range. The others are designed to have a coupling Q of $Q_c\sim 10^6$.
\newline 
We use finite-element simulations (Sonnet) to target resonator frequencies based on geometry and kinetic inductance. The simulated results allow us to confirm target kinetic inductance values. Critically coupled resonators have $Q_c$ of $\sim 10^6$, while over-coupled resonators have $Q_c\sim 10^4$. The internal quality factor $Q_i$ is measured by fitting the transmission coefficient $S_{21}$ using standard circle fit procedure introduced in reference \cite{Probst2015}, \cite{CircleFitNote}.

\begin{figure}[t]
	\includegraphics[width= 3.5 in]{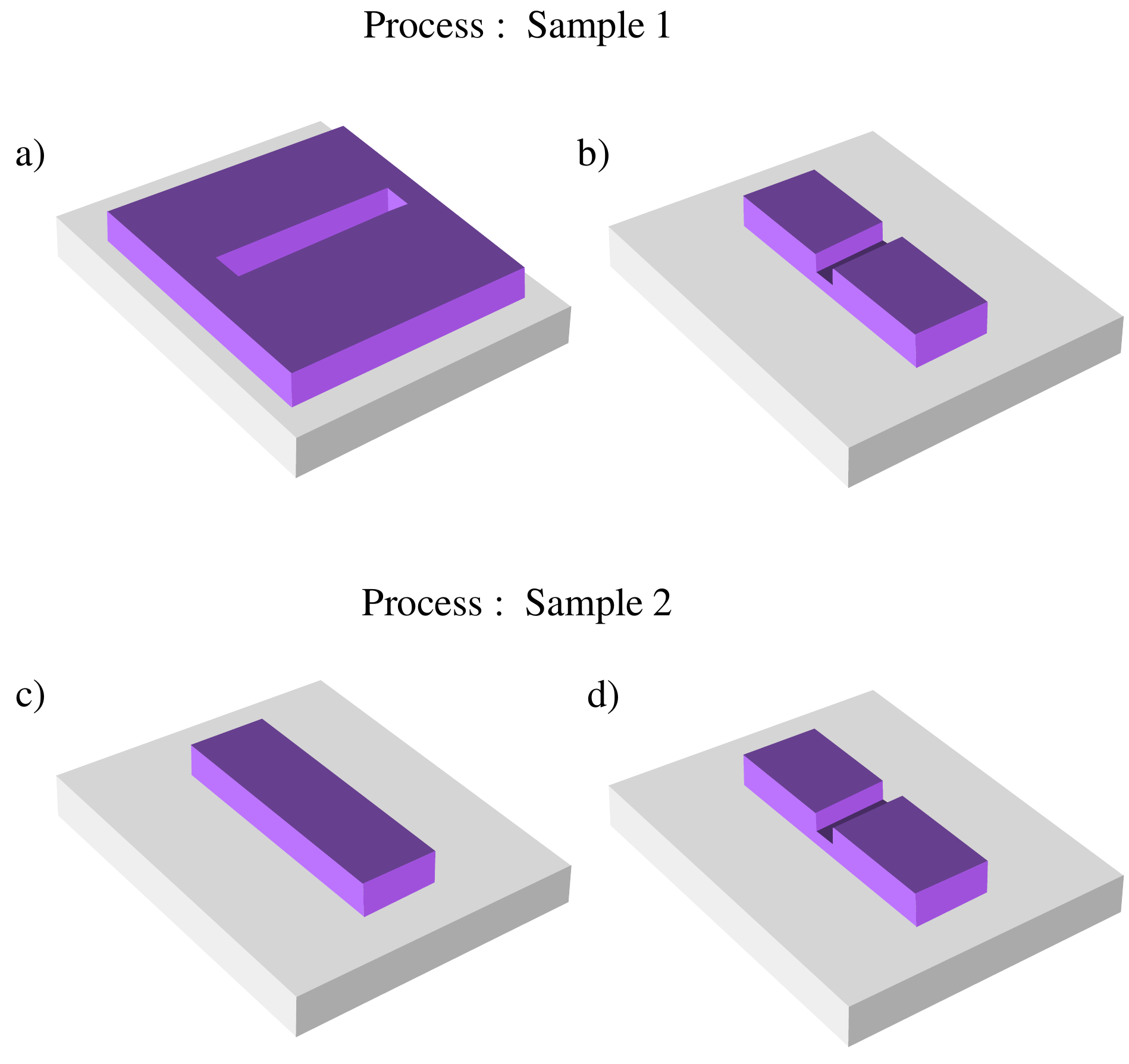}
	\caption{\textbf{Process steps for Sample 1 \& Sample 2} a, b) Fabrication process where the lead pattern is etched first and the constriction geometry for the weak link is etched second. c, d) Fabrication process where the constriction geometry for the weak link is etched first and the lead geometry is defined second.
			}
	\label{figS3}
\end{figure}
\begin{figure*}[t]
	\includegraphics[width= 5 in]{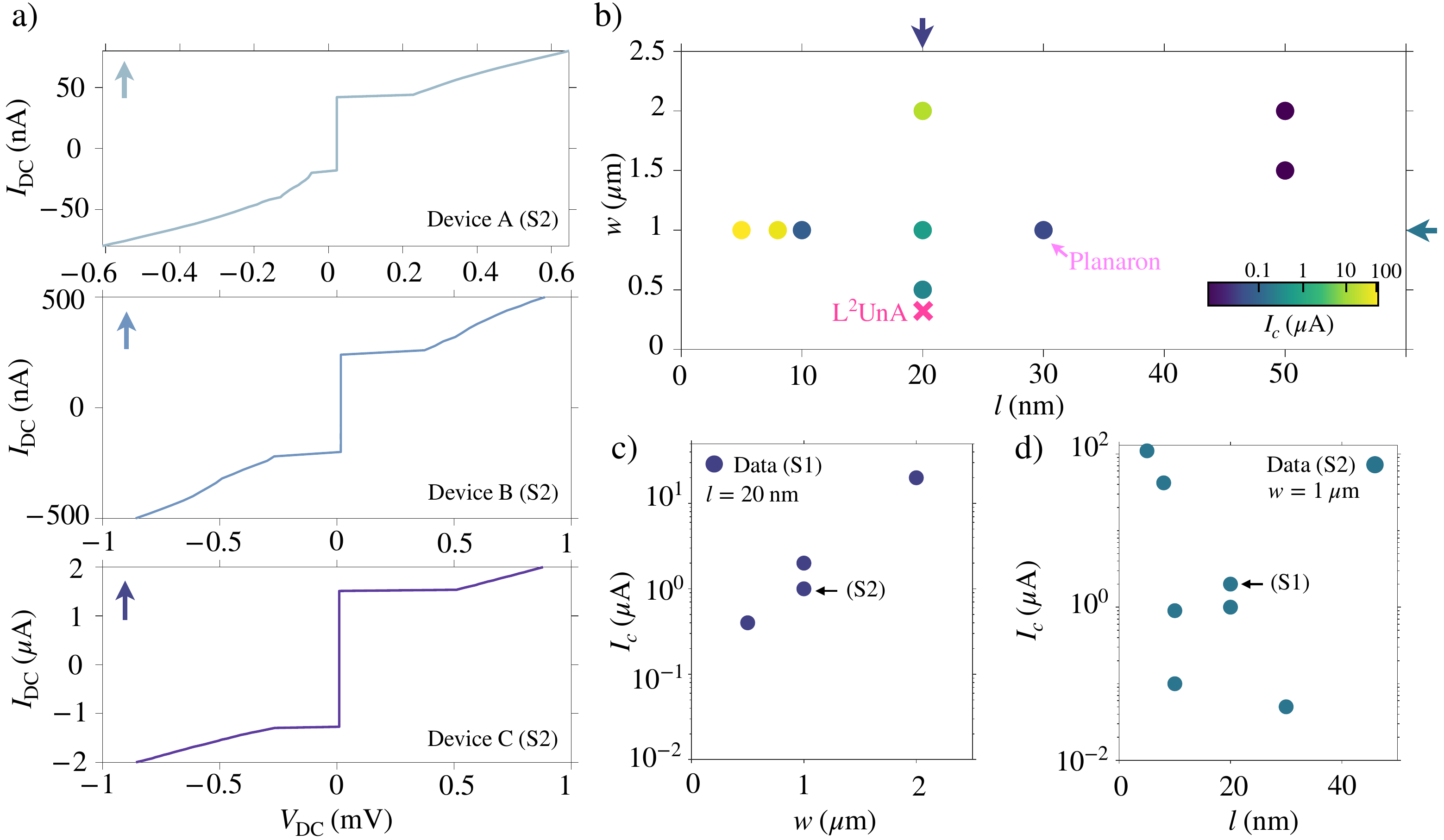}
	\caption{\textbf{Critical current measurements across devices} a) Current - voltage characteristics of three weak links from sample S2. We observe critical current, $I_c$,  values change by orders of magnitude as weak link width is varied. b) Superconducting and insulating behavior of weak links are characterized by critical current values, represented in the color bar. Data from measured samples are represented as colored circles on the color plot of critical current as a function of their weak link width and length. Data points are labeled with their respective device names. c) Critical current data as a function of weak link width from measurements done on sample S1, represented by solid purple circles. d) Critical current data as a function of weak link length from measurements done on sample S2, represented by solid teal circles.}
	\label{icscaling}
\end{figure*} 

Across all geometries, we observe internal quality factors exceeding $10^6$ at single-photon powers, including resonators with kinetic inductance values as high as $250 \:\mathrm{pH/\square}$ [Fig. \ref{figS2}a]. We find that ALE treatment plays a critical role in improving surface quality. This improvement in surface quality correlates with higher measured quality factors in CPW resonators [Fig. \ref{figS2}b]. 
\newline 
We use atomic force microscopy (AFM) to characterize surface roughness of the NbN films before and after ALE. Insets in Fig. \ref{figS2}b show AFM results on two representative samples, revealing that the ALE process reduces the root-mean-square (RMS) roughness by over 50\% compared to untreated NbN films. The origin of large surface roughness before ALE treatment remain a question for future studies.  

\section{C. Fabrication of single-film weak links: Sample 1 and Sample 2}
We explored two fabrication methods for defining NbN weak links. In the first approach, used for Sample 1 (S1), we begin by etching a trench into the uniform NbN film to define the weak link constriction [Fig.~\ref{figS3}a]. This is followed by a second lithography step and etching to define the lead geometry [Fig.~\ref{figS3}b]. In the second approach, used for Sample 2 (S2), the sequence is reversed: the leads are defined first by etching the overall lead structure, and the weak link is subsequently formed by etching a constriction across this lead rectangle [Fig.~\ref{figS3}c,d]. We speculate that this reversed sequence may expose the lead sidewalls, making them susceptible to oxidation and as a result alter local etch rates. Both fabrication methods produce devices with qualitatively similar behavior in transport measurements. However, differences emerge under applied magnetic fields, where variations in interference patterns are observed (See Section XI).

\section{D. Scaling of critical current with weak link width and length}
Here, we summarize critical current, $I_c$, measurements for both samples (S1 and S2), revealing highly tunable critical currents across devices on each sample. The measured $I_c$ range from a few tens of nA to tens of $\mu A$, controlled by weak link length ($l$) and width ($w$).
Figure~\ref{icscaling}a presents a selection of current-voltage ($I$–$V$) characteristics for three superconducting devices from sample S2. As noted in the main text, the extracted $I_cR_N$ products deviate from the zero-temperature Ambegaokar-Baratoff relation for an S-I-S junction, $I_cR_N = \pi\Delta(0)/2e$. The computed $I_cR_N$ values fall between 0.24~mV and 0.44~mV—nearly an order of magnitude lower than expected for conventional S-I-S junctions (see main text for detailed discussion).
We observe that $I_c$ decreases rapidly with both length, $l$ and width, $w$ of the weak link [Fig. ~\ref{icscaling}b and~\ref{icscaling}c]. The 2D map in Fig. ~\ref{icscaling}b also includes data on insulating weak links with $l = 50\:\mathrm{nm}$ [main text Fig. 2e-f]. Parameters used to realize a qubit (planaron) and linear resonator (Linear Lumped Unilateral Admitter, $\mathrm{L^2UnA}$) are indicated on the same plot.
\begin{figure*}[t]
    \includegraphics[width= 7 in]{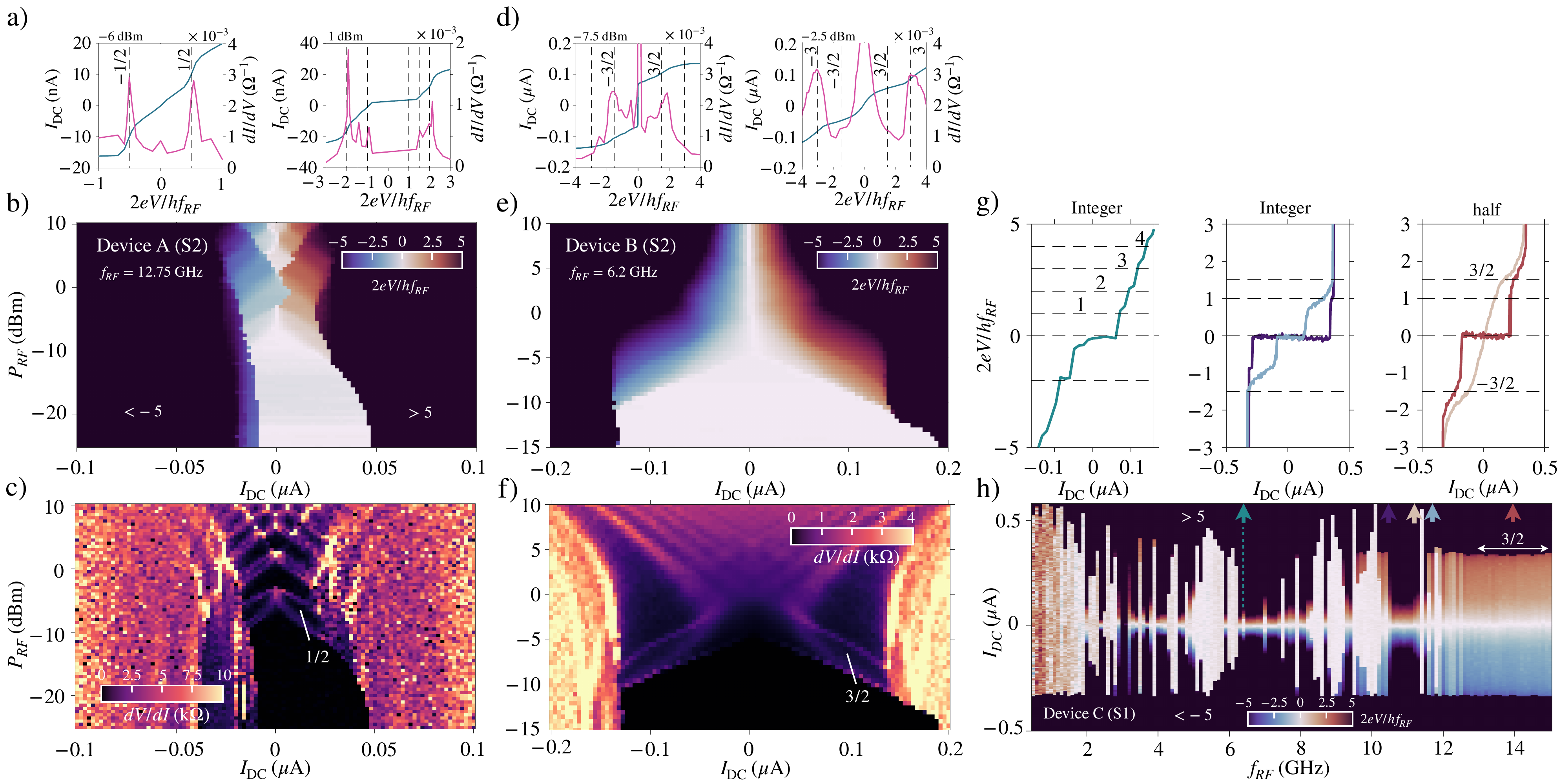} 
    \caption{\textbf{Half and integer Shapiro steps across devices} a-c) Device A (S2) measured at 12.75 GHz as a function of DC bias and RF power. a) Half-integer Shapiro steps are observed at different RF powers shown as line cuts taken from DC voltage map in panel b) and measured differential resistance $dV/dI$ map in panel c). d-f) Measurements of Device B (S2) at 6.2 GHz, showing half-integer Shapiro steps at selected RF powers. d) Line cuts taken from DC voltage map in panel e) and $dV/dI$ map in panel f). g) Measurements of Device C (S1) as a function of DC bias and RF frequency show both integer and half-integer Shapiro steps at selected RF frequencies. Colors of different lines correspond to the frequencies that are marked by the same colored arrows DC voltage map in panel h).}
    \label{shapiro-fig}
\end{figure*}

\section{E. Transport measurements: weak links}
We perform transport characterizations of the weak links in a dilution refrigerator with a base temperature of $\sim 20$ mK. The dilution refrigerator is equipped with a 3-axis magnet (6-1-1 T) used to observe Fraunhofer interference patterns (appendix K). All electrical lines are thoroughly filtered through a QDevil (RF and RC) filter, which is mounted on the mixing chamber plate. 
The differential resistance ($dV/dI=V_{\rm AC}/I_{\rm AC }$) is recorded using standard lock-in techniques with an AC frequency between $\sim 17$ and $\sim 76$ Hz: an AC bias was applied to the device input line while a separate line provides the path to ground. The four terminal AC voltage is simultaneously recorded. For superconducting devices we measure their current-voltage ($I-V$) characteristic in a constant current configuration: an AC bias, $I_{AC}$ between 0.5 nA and 5 nA is supplied with a DC bias current superimposed. The four terminal DC and AC voltage is recorded. For the lowest critical current devices we use the smallest AC bias current to prevent heating. For insulating weak links, we bias with a constant voltage configuration using $50 \: \mu\rm V$. Nonlinear $I-V$  was measured by sweeping a DC voltage on top of the AC voltage and recording the injected DC current with the four terminal AC and DC voltages.
    
\section{F. Half and integer Shapiro steps across samples and devices}
Additional data summarizing half-integer Shapiro steps in several representative devices are shown. Figure ~\ref{shapiro-fig}a,b include DC voltage and differential resistance, $dV/dI$, maps for Device A (S2) [main text Fig.~2a,b] and Device B (S2) [main text Fig.~2c,d], both of which show strong evidence of half-integer steps. Note that the RF frequencies used here differ from those used in the main text. These frequencies were chosen because they yield the clearest half-step signatures in the respective devices. Figure~\ref{shapiro-fig}c presents a broad DC voltage map with a selection of line cuts for another device as a function of RF frequency. We observe transitions from integer to half-integer voltage steps, with the half-integer steps becoming more prominent at higher frequencies.
\newline
Finally, we also observe missing Shapiro steps with a strong 3/2 step across devices and frequencies [Fig.~\ref{shapiro-fig}], which is currently not well understood. Missing Shapiro steps may arise from different origins, such as trivial effects including Landau-Zener transitions (LZT) of high-transparency modes in multi-mode junctions \cite{dartiailh2021}, and in systems (not including these results) where intrinsic topological properties may be present \cite{wiedenmann2016}. In the latter, exclusively missing \textit{odd} Shapiro steps is expected.

\section{G. Planaron Fabrication}
The planaron qubit is fabricated from a single NbN film. All structures are defined via electron-beam lithography followed by etching with ALE. 
First, we define a trench by exposing and etching a narrow constriction into the NbN film (similar to S1 processing, appendix C.). This creates a localized weak link that will result in the nonlinear element of the qubit. After defining the weak link, the capacitor pads and large-scale lead geometry are patterned and etched to complete the transmon qubit layout. All etching is performed selectively on the NbN film through ALE and the sample is then cleaned in buffered oxide etch (BOE 6:1) for an extended period ($\sim20$ min). BOE does not attack NbN, allowing for effective removal of residual contaminants, organics and native oxides without damaging the device.

\section{H. Planaron Parameters}
We have performed a two-tone spectroscopy on our planaron qubit to quantify qubit parameters including the anharmonicity (see main text). We can compare these results to the design parameters of our planaron qubit. Conventional transmon parameters were used. A shunt capacitance of $C_S=68.7 \:\mathrm{fF}$ extracted from finite-element simulations, corresponding to a charging energy of $E_C/\hbar=2\pi\times293\:\mathrm{MHz} $, and computed anharmonicity is $\alpha= 2\pi\times 306\:\mathrm{MHz}$, where an extra $\sim4\% $ comes from the numerical diagonalization \cite{koch2007}. We extract a geometric and kinetic stray inductance from finite-element simulations including the sheet inductance of NbN of $L_K=17.5\:\mathrm{pH/\square}$, resulting in a total stray inductance of $L_S\sim 0.2\:\mathrm{nH}$. The junction provides an estimated Josephson energy of $E_J/\hbar =2\pi \times 26.15\:\mathrm{GHz}$, using that $\hbar\omega_q=\sqrt{8E_JE_C}-E_C$,
which corresponds to a Josephson inductance of $L_J=6.25\:\mathrm{nH}$, that places the planaron in the transmon regime with $E_J/E_C\sim90$. 
Our results show a reduced measured anharmonicity, which cannot be explained only by stray kinetic inductance, and is therefore more likely explained by microscopics of NbN weak link, see main text. Recently, higher harmonics in the current-phase relation of Al/AlOx/Al Josephson junctions have also been measured \cite{willsch2024}.
\begin{figure}[t]
    \includegraphics[width= 3.5 in]{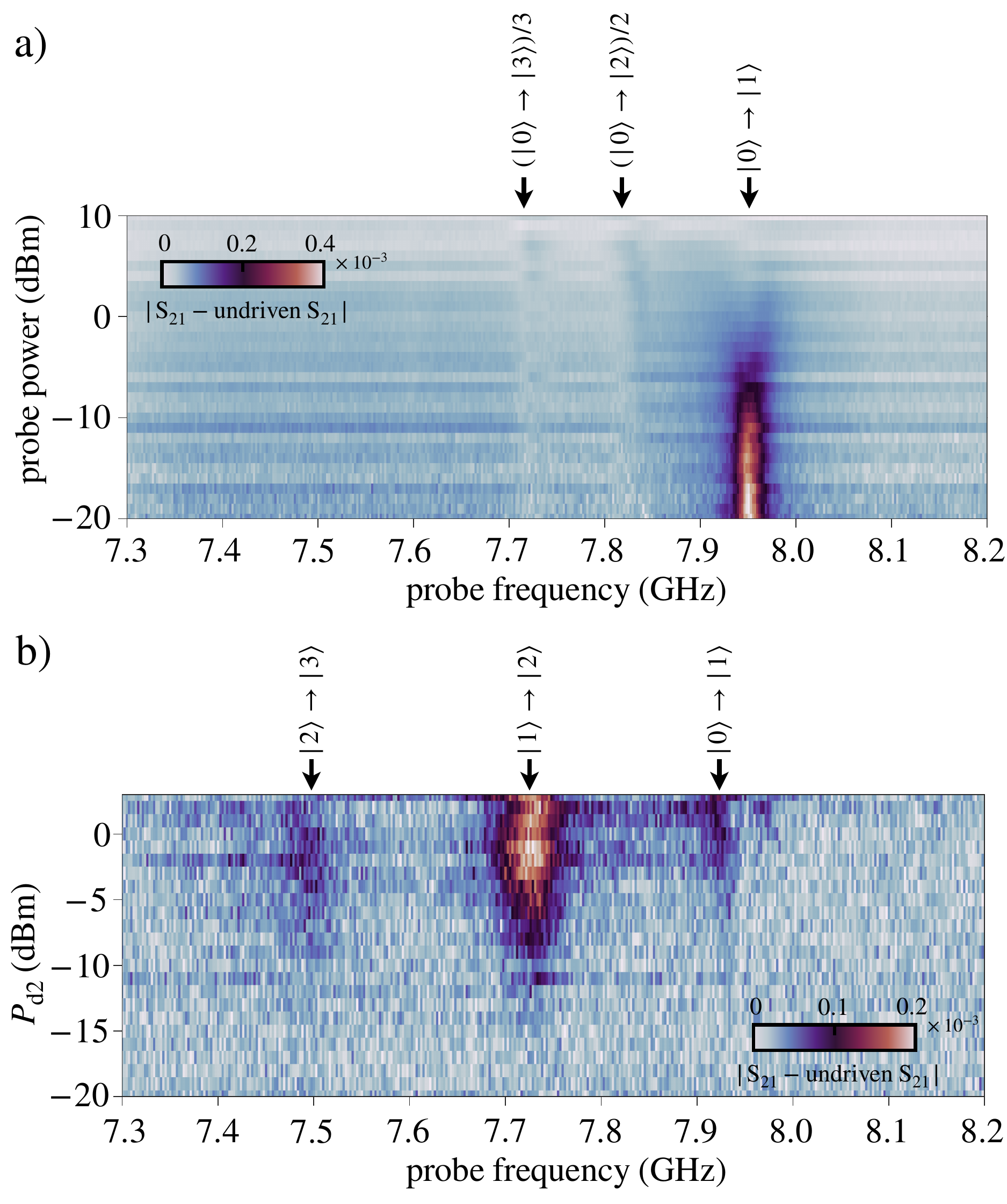} 
    \caption{\textbf{Planaron spectroscopy} Additional spectroscopy data for the planaron qubit. a) Single-tone spectroscopy as a function of power. The high-power response of the device is subtracted from this plot. The bright feature at $7.95 \mathrm{\ GHz}$ is the $|0\rangle \rightarrow|1\rangle$ transition. The fainter line at $7.83 \mathrm{\ GHz}$ can be attributed to a $|0\rangle \rightarrow|2\rangle$ transition using two drive photons, and the line at $7.72 \mathrm{\ GHz}$ is consistent with $|0\rangle \rightarrow|3\rangle$ using three drive photons. b) Three-tone spectroscopy. While driving at $\omega_{01}$ and $\omega_{12}$ simultaneously, we sweep the frequency and power of a third spectroscopy tone. We subtract the low-power response of the system in this configuration from the plot. In addition to seeing the $|0\rangle \rightarrow|1\rangle$ and $|1\rangle \rightarrow|2\rangle$ transitions, a third line appears at $7.48 \mathrm{\ GHz}$, which we attribute to the presence of the $|2\rangle \rightarrow|3\rangle$ transition.}
    \label{more spec}
\end{figure}
\section{I. Additional Spectroscopy data}
In addition to the data shown in the main text, we also identify the $|2\rangle \rightarrow|3\rangle$ transition of the device using 3-tone spectroscopy, see Fig \ref{more spec} b. The $|2\rangle \rightarrow|3\rangle$ transition is $235 \mathrm{\ MHz}$ below the $|1\rangle \rightarrow|2\rangle$ transition. Based on the spectrum of the device, we could in principle extract additional information about the energy-phase relation of the weak link. However, the strong dissipation in our device limits the number of visible transitions, so we leave this to future work.

We also show extended single-tone spectroscopy data at high power in Fig \ref{more spec} a. As in the data shown in the main text, the high-power transmission is subtracted from the data. The $|0\rangle \rightarrow|1\rangle$ transition is the brightest line in this dataset. There are two other transitions visible in this plot. We attribute one to the $|0\rangle \rightarrow|2\rangle$ transition using two drive photons. The frequency of this transition, at $(\omega_{01}+\omega_{12})/2$, supports this hypothesis. A third line, which we suspect to be a $|0\rangle \rightarrow|3\rangle$ transition using three drive photons, and appears at about $(\omega_{01}+\omega_{12} + \omega_{23})/3$, where these frequencies are determined from the three-tone spectroscopy. 

\begin{figure}[t]
	\includegraphics[width= 3.5 in]{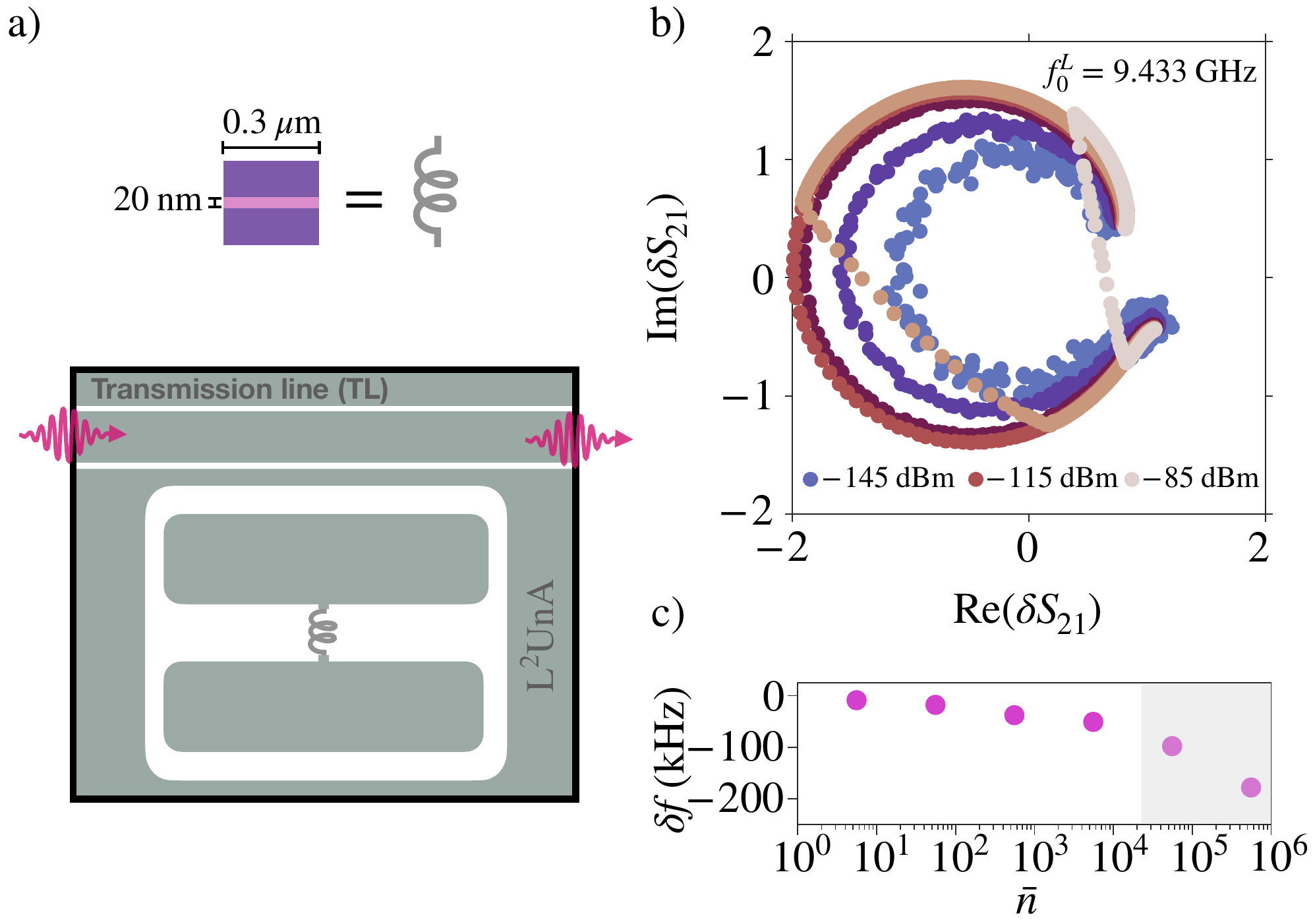}
	\caption{ \textbf{Single-tone spectroscopy of linear resonator} Single-tone spectroscopy data for a device with different weak link geometry than planaron qubit. We use a linear inductor symbol to distinguish the butterfly weak links employed in planaron qubit (see main text). a) Dimensions of the etched region and device layout. b) Power dependence of the lineshape. Unlike the response of the device shown in the main text, the lineshape of this device remains a circle until a critical power where the phaseroll "skips" a portion of the circle. This response is indicative of an oscillator with anharmonicity smaller than the linewidth. c) Resonant frequency as a function of drive power. We extract an anharmonicity $\alpha/2\pi \approx 10 \mathrm{\ Hz}$, much less than the charging energy $E_C/h\approx 300 \mathrm{\ MHz}$.
			}
	\label{linmon}
\end{figure}

\section{J. Microwave response of additional devices}
We also measured the microwave response of devices on a second chip. These devices had a different weak link geometry. An example is shown in Figure \ref{linmon}. Compared to the planaron, this device has similar capacitance but a reduced weak link length of $20 \mathrm{\ nm}$ and a reduced width of $0.3 \mathrm{\ \mu m}$. We measure several devices with resonant frequencies around $9.4 \mathrm{\ GHz}$, suggesting that their inductance is similar to that of the planaron. However, the nonlinearity of these devices is orders of magnitude smaller. We estimate $\alpha/2\pi\approx 10 \mathrm{\ Hz}$. We leave the investigation of how nonlinearity depends on weak link geometry to future work. We determine the anharmonicity of these samples based on the shift in resonance frequency as a function of drive power; this is only possible for devices with $\alpha \ll \kappa_t$. In this limit, the lineshape of the weakly anharmonic resonator remains circular until a critical photon number is reached, when a portion of the phase roll gets "skipped" due to bifurcation. The size of the resonant circle changes with the drive power due to a photon number-dependent quality factor, but the shape does not become "squashed", as it does in devices with $\alpha \gg \kappa_t$, such as the planaron shown in the main text. Notably, the devices displaying low anharmonicity had a much higher sinlgle-photon internal quality factor $Q_i = 3.8\times10^5$. This puts them in a potentially useful regime of being highly lumped with high quality factor and extremely small anharmonicity.

\begin{figure}[t]
	\includegraphics[width= 3.5 in]{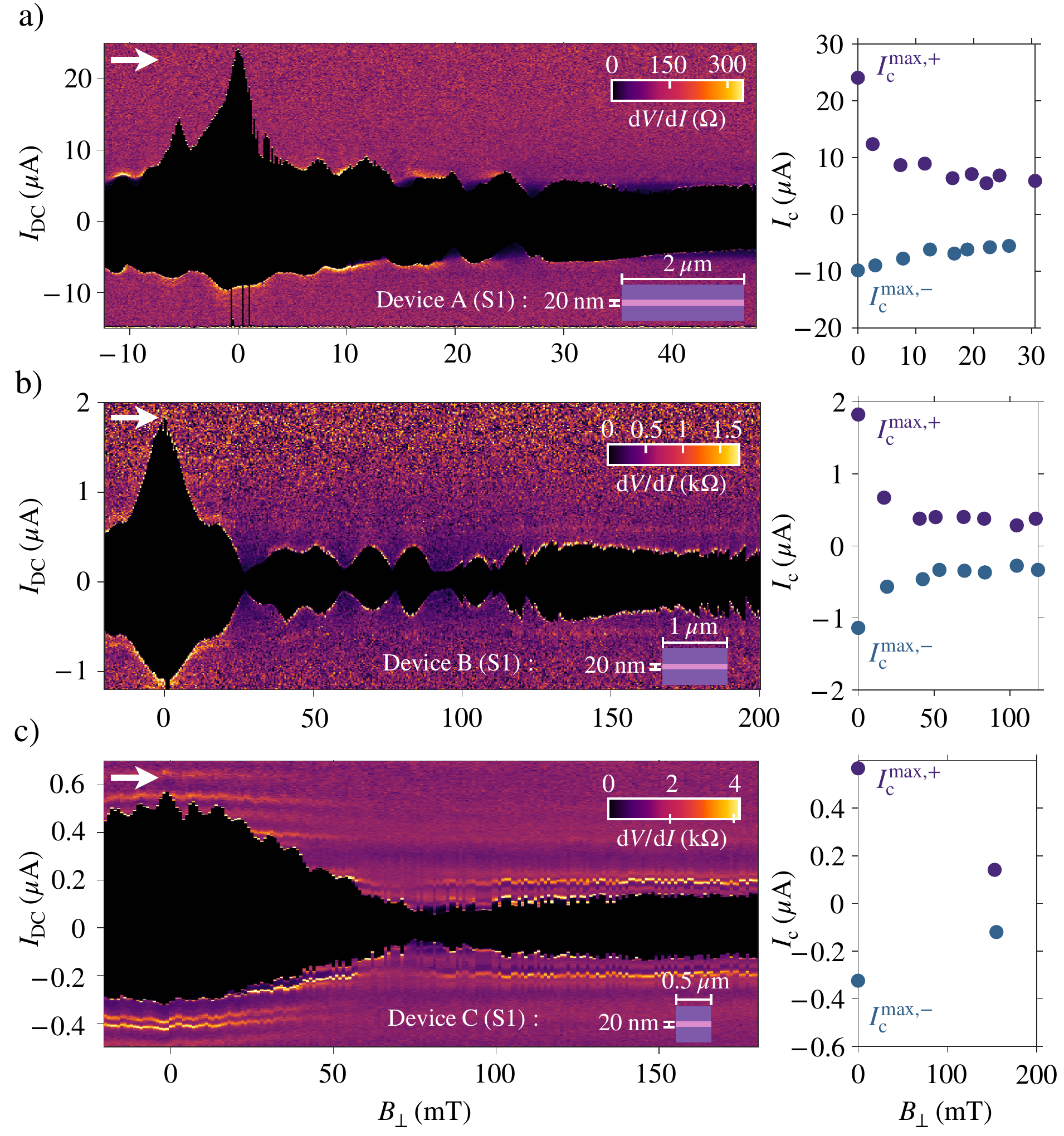}
	\caption{\textbf{Single-slit interference pattern} a-c) Magnetic field dependence of weak links fabricated with method referred to as S1. Each device has a different width a) $2\: \mu m$, b) $1\: \mu m$ and c) $0.5\: \mu m$ but similar length of 20 nm. Each panel shows both the measured Fraunhofer interference pattern (left) and the extracted max critical current $I^{\rm max,+/-}_c$ (right) as a function of applied field. The observed decrease in the magnitude of critical current with each lobe is consistent with a typical single-slit interference pattern. 
			}
	\label{fraunhofer}
\end{figure}

\section{K. Magnetic flux dependence}
\indent External flux applied to a Josephson junction can give insights into the supercurrent flow. The supercurrent can be modulated by an applied perpendicular field, $B$, with period modulation corresponding to one flux quantum through the weak link, $B_0=\Phi_0/A$, where $\Phi_0=h/2e$ is the magnetic flux quantum and area of the weak link is given by $A=(l+2\lambda)w$, where $\lambda$ is the London penetration length. As noted previously, the sequence in which the weak link and lead is fabricated influences the supercurrent profile (appendix C.).

\begin{figure}[t]
	\includegraphics[width= 3.5 in]{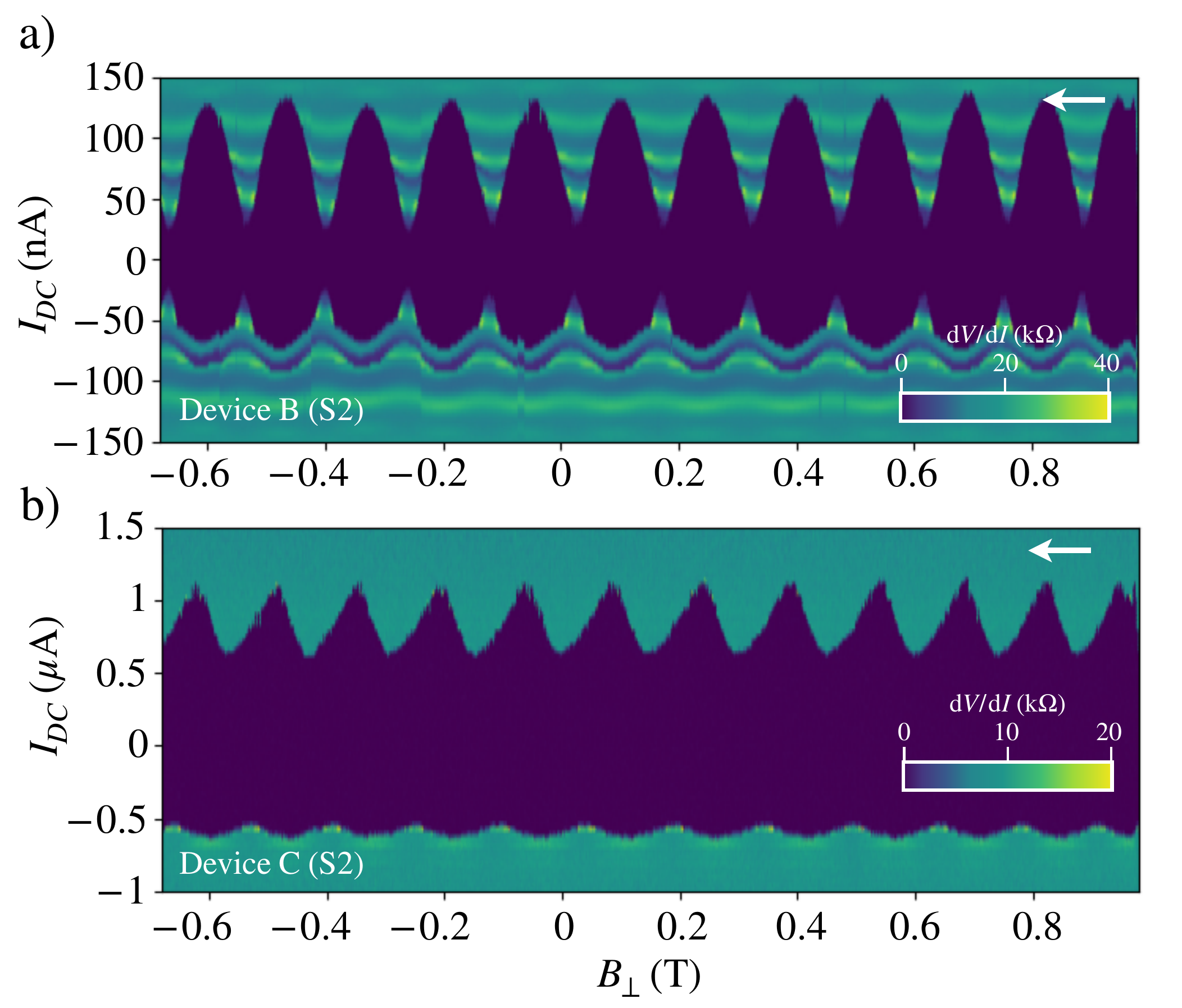}
	\caption{ \textbf{Double-slit interference pattern} a, b) Differential resistance $dV/dI$ maps as a function of applied magnetic field (horizontal axis) and DC bias current (vertical axis) for two weak links fabricated with method referred to as S2. The interference patterns reveal periodic suppression of the supercurrent with applied flux. The lobes have equal heights characteristic of a double-slit interference pattern, suggesting that the supercurrent primarily flows along the edges of the weak link.
			}
	\label{fraunhofer2}
\end{figure}

\subsection{Sample 1}
A homogeneous supercurrent flow gives rise to a  single-slit Fraunhofer interference pattern characterized by a central lobe with height $I^{max}_c$ and side lobes decaying as $1/B$. A map of the differential resistance as a function of the applied perpendicular field, $B_\perp$, and bias current, $I_{DC}$, is shown in Fig.~\ref{fraunhofer} for three devices on S1. Each device have a different area and critical current defined by width of the weak link ($w=\{2,\:1,\: 0.5 \}\:\mu m$), keeping a fixed length ($l=20\:\mathrm{nm}$). It is evident that the measured interference pattern behave qualitatively as expected with a central lobe and decaying side lobes. This type of interference pattern suggests a predominantly homogeneous current flow through the weak link, although a few characteristics deviate from a typical Fraunhofer pattern. We observe a non-zero $I_c$ at the nodes, which could arise from different types of inhomogeneity that leads to either asymmetric current flow \cite{Dynes1971}, or stemming from nonuniform weak link boundaries, both of which could be a result of the fabrication process or related to microscopics in the NbN barrier. To obtain a quantitative analysis we use the interference envelope to extract $I^{\mathrm{max+}}_c(B),I^{\mathrm{max-}}_c(B)$, shown as side panels in Fig.~\ref{fraunhofer} and plotted against $B$. Due to strong hysteresis in the weak links we find $I^{\mathrm{max+}}_c(B)\neq I^{\mathrm{max-}}$. Current is swept from negative DC bias to positive in Fig.~\ref{fraunhofer}.
\newline
We further estimate modulation period for each weak link, to be $B^{\mathrm{exp.}}_0=\{3.3, \:16.7,\:70\}\:\mathrm{mT}$, showing an increase with decreased weak link area as expected, although deviate from the estimated periods $B^{\mathrm{theory}}_0=\{2.4,\:4.9,\:9.8\}\:\mathrm{mT}$, taking $\lambda_{\mathrm{NbN}}=200\:\mathrm{nm}$. This could be due simple overestimation of the effective weak link area or imperfect screening from the leads; in particular, when the width of the weak link is comparable to the penetration depth of NbN, there is the largest discrepancy in modulation period [Fig.~\ref{fraunhofer}c].\newline

\subsection{Sample 2}
Modulations in the critical current as a function of applied magnetic field display lobes with similar height consistent with a double-slit interference patterns, primarily observed for weak links on S2, as illustrated in Fig.~\ref{fraunhofer2}. This suggests that the supercurrent is not distributed uniformly across the weak link but is instead concentrated near the edges. This edge-dominated transport indicates that the central region of the weak link may have suppressed transport due to the fabrication process (detailed in appendix C.). Notably, we also observe lifted nodes in S2 i.e. non-zero critical current similar to S1, which is attributed to inhomogeneity in the barrier or nonuniform boundaries.



%% file: manuscript.bbl
\begin{thebibliography}{10}
\bibitem{CircleFitNote}
{The version of the circle fitting algorithm used in this paper has been modified by Chan U. Lee, Suhas Ganjam and Yanhao Wang }.
\end{thebibliography}
